\documentstyle[aps,prl,eqsecnum,tighten]{revtex}
\begin{document} \title {
Crossover and scaling in a nearly antiferromagnetic Fermi liquid in two
dimensions}
\author{Subir Sachdev}
\address{Department of Physics, P.O.
Box 208120, \\ Yale University, New Haven, CT 06520-8120}
\author{Andrey V. Chubukov}
\address{
Department of Physics, University of Wisconsin, Madison, WI 53706\\
and P.L. Kapitza Institute for
Physical Problems, Moscow, Russia}
\author{Alexander Sokol}
\address{Department of Physics, University of Illinois at Urbana-Champaign,
Urbana IL 61801\\
and L.D. Landau Institute for Theoretical Physics, Moscow, Russia}
\date{\today}
\maketitle

\begin{abstract}
We consider two-dimensional Fermi liquids in the vicinity of a quantum
transition
to a phase with commensurate, antiferromagnetic long-range order.
Depending upon the Fermi surface topology,
mean-field spin-density-wave theory predicts two different types
of such transitions, with {\em mean-field\/} dynamic critical exponents
$z=1$ (when the Fermi surface does not cross the magnetic zone boundary,
type $A$) and $z=2$ (when the Fermi surface crosses the magnetic zone
boundary, type $B$).
The type $A$ system only displays $z=1$
behavior at all energies and its scaling properties are
similar (though not identical) to those of an insulating Heisenberg
antiferromagnet.
Under suitable
conditions precisely stated in this paper, the type $B$ system displays a
crossover from relaxational behavior at low energies to type $A$
behavior at high energies.
A scaling hypothesis is proposed to describe this crossover:
we postulate a universal scaling function which
determines
the entire,
temperature-, wavevector-, and frequency-dependent, dynamic,
staggered spin susceptibility
in terms of 4 measurable, $T=0$,
parameters (determining the distance, energy, and order parameter
scales, plus one crossover parameter).
The scaling function contains the full scaling behavior
in all regimes for both type $A$ and $B$ systems.
The crossover behavior of the uniform susceptibility and the specific heat
is somewhat more
complicated and is also discussed.
Explicit computation of the crossover functions
is carried out in a large $N$ expansion on a mean-field model.
Some new results for the
critical properties on the ordered side
of the transition are also obtained in
a spin-density wave formalism. The possible relevance of our results to the
doped
cuprate compounds is briefly discussed.
\end{abstract}

\tableofcontents

\section{Introduction}
\label{intro}
A number of recent works~\cite{CHN,tsvelik,jinwu,Ch-Sach,Millis,Sokol,pines,su}
have proposed a description
of the low temperature spin dynamics of strongly correlated electronic
systems using their proximity
to an actual or hypothetical zero temperature, magnetic, quantum phase
transition. Such an approach
has the advantage of allowing development of a systematic expansion about a
point with non-trivial
spin correlations and with strongly interacting excitations.
Some of these proposals have arisen in the context of
two-dimensional interacting
electron models of the doped cuprate superconductors,
while others considered the heavy-fermion compounds. The ideas of this
paper will be presented in the
former context, although our approach
may be of a more
general utility.

We begin our discussion by reviewing two recent complementary approaches to
magnetic quantum transitions in the
cuprate compounds. A unified picture of spin fluctuations in
the
lightly-doped cuprates obtained from these past approaches, and our present
work,
is provided towards the end of this paper in Sec.~\ref{conc}.
We wish to emphasize at the outset that all of the discussion in this paper
refers
to spin fluctuations associated with {\em commensurate\/} antiferromagnetic
ordering;
for the cuprates this corresponds to an ordering wavevector $Q= (\pi , \pi)$.
We have little to say here about the case of incommensurate ordering.

The first approach begins from the insulating parent compound, like $La_2
Cu O_4$, whose spin
fluctuations can be modeled by
a spin-1/2 Heisenberg antiferromagnet on a square
lattice~\cite{CHN}.
The low energy excitations of the antiferromagnet are believed to be
well-described by a
continuum ${\rm O}(3)$ non-linear sigma model field theory. The non-linear
sigma
model is parameterized by a single coupling constant, $g$, which measures the
strength of quantum fluctuations in the system; for $g < g_c$, the
system has N\'{e}el order, while for $g > g_c$, it is in the quantum disordered
phase. The transition at $g = g_c$ has been
studied in some detail~\cite{Ch-Sach}---it has dynamic critical exponent
$z=1$ and leads to a quantum disordered phase in which at $T=0$ the low energy
magnon excitations
have a gap and an infinite lifetime. The current experimental and
theoretical consensus is that the
$S=1/2$ Heisenberg antiferromagnet has
macroscopic Neel order in the
ground state, and therefore should map onto a sigma model with $g<g_c$.
Now consider doping this antiferromagnet with a small number of holes. At very
small doping, the holes form small elliptical pockets at $(\pi/2,\pi/2)$
and symmetry related
points in the magnetic Brillouin zone~\cite{shraiman,chub-fren,hole-pockets}.
Quantum fluctuations associated with these holes decrease the spin
stiffness of the
antiferromagnet~\cite{shraiman} and should therefore drive the effective
value of $g$ closer to
the quantum transition point at $g=g_c$. (We are assuming here that
conditions are such that
there is a direct transition from a commensurate long-range ordered state
to a commensurate
quantum-disordered state with increasing doping~\cite{sss}; we are
neglecting the possibility
of an intermediate incommensurate long-range ordered state~\cite{ShSi,karen}).
 Suppose that quantum fluctuations are strong enough
that the disordering transition occurs while the
holes still occupy small pockets. The critical properties of such a transition
were studied by one of us~\cite{sss} in the framework of
the Shraiman-Siggia model~\cite{ShSi}: it was found that the dynamic
critical exponent remained
at $z=1$, and the mobile charge carriers
only introduced a small, but negligible damping of the magnon excitations at
the critical point.
The full structure of the quantum disordered phase in this model, and in
particular, the topology of
its Fermi surface is not well understood: we will discuss
these issues further in
Sec~\ref{secss}.

The second approach begins with the opposite limit of large doping, where
the dilute system of electrons
is presumably well described as a Fermi liquid.
Using mean-field spin-density-wave ideas, a scenario for the onset of
antiferromagnetic
order in such a Fermi liquid was proposed many years ago by
Hertz~\cite{hertz}, and extended recently by Millis~\cite{Millis}.
It turns out to be important to distinguish two cases~\cite{Millis}
depending upon the value
the ordering wavevector  $Q$ (for our case $Q=(\pi, \pi)$),
 and the shape of the
Fermi surface in the
quantum-disordered phase.\\
{\em (A) Damping of spin excitations with momenta near $Q$,
due to conversion into particle-hole pairs, is forbidden:}
This is the case when
$Q$ cannot connect two points on the Fermi surface. For a
circular Fermi surface
this corresponds to $Q > 2 k_F$ where $k_F$ is the Fermi wavevector.
The transition then has the mean-field exponent
$z=1$ and is rather similar to the discussion above on the
first approach. The scaling results of Ref.~\cite{Ch-Sach} apply mostly
unchanged:
some simple modifications are necessary for the uniform susceptibility,
and are discussed in Sec.~\ref{secunisus}
 \\
{\em (B) Damping of spin excitations with momenta near $Q$,
due to conversion into particle-hole pairs, is allowed:} This
is the case when $Q$ can connect two points on the Fermi surface,
which for a circular Fermi surface corresponds to $Q < 2 k_F$.
This transition has
$z=2$ in mean-field theory. Accordingly, the magnon excitations in the
quantum disordered
phase are overdamped and relaxational.

We will not discuss the special case  $Q =2k_F$ in
this paper.

A phenomenological form for the magnetic susceptibility near a type $B$
transition was
introduced in context of cuprate superconductors by Millis, Monien and
Pines~\cite{MMP} to explain NMR data in $YBa_2Cu_3O_7$.
The $z=1$ to $z=2$ crossover with doping was later proposed by Pines and
one of us~\cite{Sokol} and by Barzykin {\em et al.}~\cite{pines}
to describe evolution of the experimental NMR and neutron scattering data
with oxygen concentration for $YBa_2Cu_3O_{6+x}$.
 Very
recently, Liu and Su~\cite{su} considered
a two-component ``Kondo-lattice''-type
model of the cuprates, with
separate localized spin and itinerant electron degrees of freedom. They
assumed conditions that were appropriate to have a type $B$ phase transition
with $z=2$,
and found
behavior characteristic of this value of $z$ at the lowest
energies or temperatures.
At higher temperatures however, they found a crossover to behavior
characteristic of $z=1$. Earlier,
the temperature-induced crossover to the $z=1$ regime was
discussed in Ref \cite{Sokol:LANL} in relation
to $YBa_2Cu_3O_{6.63}$.
It is not difficult to see that this crossover is in fact a rather general
phenomenon for type B systems ---
there will always be reactive terms
$\sim \omega^2$ ($\omega$ is a measuring frequency)
in spin response functions which
will overwhelm dissipative terms at large enough
$\omega$. The crossover should be especially pronounced in systems
where the damping constant, $\gamma$, is small.

The main purpose of this paper is to discuss the scaling properties
of the quantum disordered phase of systems near transitions of type $B$.
We will also make a few comments about type $A$ systems to which
the analysis in Ref.~\cite{Ch-Sach} will mostly apply.
 We will assume that
conditions in the type $B$ system are
such that the higher energy crossover to type $A$ behavior occurs at energies
which are significantly smaller
than other high energy cutoffs like the Fermi energy or the exchange
constant. (For type $B$ systems which violate this condition, our results
reproduce the correct asymptotic critical singularities, but make a
particular choice
for numerical scale-factors which are in fact non-universal in this case;
these statements will be made more precise later).
We also assume, of course, that no other unrelated low energy scale appears.
 We will propose a scaling hypothesis, in which  the wavevector ($q$),
frequency ($\omega$),
and temperature ($T$) dependence of
the staggered spin susceptibility can be expressed in terms
of a scaling function which
involves only four experimentally determinable, $T=0$
input parameters. Only
these four parameters are dependent on the details of the microscopic
interactions in the ground state; everything else  is
universal and can, in principle, be computed in a long distance field
theory. The general scaling arguments will be presented in the
Sec~\ref{secscalinganalysis}.
The crossover behavior of the uniform susceptibility and the specific heat
is somewhat more
complicated and requires additional microscopic parameters---this is
discussed in
Secs~\ref{secunisus} and~\ref{secspecificheat}.

We will then illustrate these scaling ideas in two model calculations.

In Section~\ref{sdw}
 we consider magnetic phase transitions in a standard
spin-density-wave (SDW) formalism. We will examine a simple model---a
Hubbard model with first ($t$)
and second neighbor hopping ($t^{\prime}$) for the fermions, which
displays a transition of either type $A$ or type $B$,
depending on the ratio $t/t'$. Note that in both cases, the Fermi
surface in the quantum-disordered phase is large
({\em i.e.\/} encloses a volume
given by the total number of electrons).
 The SDW results for the magnetic
susceptibility
will illustrate the important differences in the nature of the spin
excitations between type $A$ and $B$
transitions.
The SDW theory will also be used to
obtain new
mean-field results
for critical properties on the magnetically ordered side of the transition.
Finally, we will briefly discuss the  relation between the SDW
results and those for the fluctuation-driven magnetic transition within
the Shraiman-Siggia model.

In Section~\ref{largeN}, we first use the results of Section~\ref{sdw} to
motivate a model field theory, $S$,
to describe the quantum-disordered phase in case $B$ and its crossover to
type $A$ behavior.
The model $S$, which relies on a mean-field approximation for the fermions,
will turn out to be precisely the one proposed by
Liu and Su~\cite{su}; we caution however that the validity of $S$ as a
description of the underlying fermionic excitations has not been
conclusively established asymptotically close to the critical point. We
will use a
large $N$ expansion to compute explicit results for
the crossover scaling functions of $S$.

Finally in Section~\ref{conc} we will discuss the experimental relevance of
our results and state
our main conclusions.

\section{Scaling hypotheses}
\label{secscalinganalysis}

We now present some general scaling ideas to describe the crossover
between type $A$ and type $B$ transitions.
In the terminology of the well-developed theory of crossover phenomena
between two critical points
in classical phase
transitions~\cite{crossover,nelson}, we need to distinguish between the primary
and secondary fixed points:
the primary fixed point has 2 relevant directions, while the secondary
point has only a single,
relevant direction associated with the ``thermal'' operator which drives
one across the transition.
In our case, it is clear that the primary fixed point is the type $A$ fixed
point with
$z=1$. The mean-field value of $z=1$ for type $A$ is expected to be robust
and not suffer any
corrections from fluctuations, as the coupling between the
spin-fluctuations and fermionic
quasiparticles is quite weak in this case.
Adding any damping to the magnon excitations should
be a relevant perturbation at this fixed point as it introduces additional low
energy
excitations
(it could be dangerously irrelevant, a possibility which we shall
briefly refer to later, but
not explore in any detail). The coupling, analogous to $g$, which tunes
the system across the transition
is the other relevant perturbation.
The type $B$ fixed
point must therefore
be secondary.
(We are also assuming here that the effect of the damping is associated
with only a {\em single\/}
relevant operator; there could be more than one. It should be easy to
extend the following scaling
analysis to this case, but we shall not do it in the interests of simplicity).
An important feature of
the standard crossover theory~\cite{crossover,nelson} is that
the crossover scaling functions between the two fixed
points are expressed in
terms of eigenoperators and exponents of the primary fixed point, while the
critical
singularities of the
secondary fixed point appear as nonanalytic behavior in the crossover
functions themselves. This result forms the basis of our analysis below.

We will restrict our scaling results below to the
to the quantum disordered side of the transition. We begin our discussion
by introducing the
four parameters which will characterize the
quantum-disordered ground
state; the finite temperature properties of the staggered spin
susceptibility, $\chi_s$,
will then be described by
universal scaling functions
of these four parameters only.
Imagine we have available,
either through experiments or computer simulations, the $T=0$ value of the
imaginary part of the local, on-site, dynamic spin
susceptibility $\chi^{\prime\prime}_L ( \omega )$ of the system of interest;
the local susceptibility is obtained by integrating
the dynamic susceptibility over momenta in the vicinity of $Q$ (where it equals
$\chi_s ( q, \omega)$).
In an insulating antiferromagnet, whose transition is described
completely by the primary, $z=1$ fixed point, $\chi_{L}^{\prime\prime}$
would have the form shown
in Fig~\ref{fig1}a~\cite{Ch-Sach}:
there is gap below which the spectral density is strictly zero, a
discontinuity at
the gap, and for large
$\omega$ we have $\chi_{L}^{\prime\prime} \sim \omega^{\eta}$ where $\eta$
is an anomalous dimension
of the $z=1$ fixed point~\cite{jinwu,Ch-Sach}. Here, by large $\omega$, we
mean frequencies which are
large compared to the spin fluctuation energy scale, but small compared to
upper cutoffs like the an
exchange constant or the Fermi energy; this and similar restrictions will
be implicitly assumed in the
remainder of  the paper.
Let us now examine the change in the spectrum due to a relevant
perturbation which
moves the system towards the secondary, type $B$, fixed point. Under
appropriate
conditions, mobile
fermionic
carriers can act as such a perturbation,
and the  damping due to
the particle-hole continuum
will introduce some sub-gap absorption:
$\chi_{L}^{\prime\prime}$
will then look like Fig~\ref{fig1}b. The gap has turned into a pseudo-gap, and
$\chi_{L}^{\prime\prime} \sim \omega $ for small $\omega$.
However, we will still have $\chi_{L}^{\prime\prime} \sim
\omega^{\eta}$ for large
$\omega$ as the primary fixed point behavior is expected to continue to
hold at large
$\omega$.
We now extract three
parameters from this form for $\chi_{L}^{\prime\prime}$:\\
({\em i\/}) An overall amplitude ${\cal Z}$: This is defined by
\begin{equation}
{\cal Z} = 4 \lim_{\omega \rightarrow \infty} \frac{\chi_{L}^{\prime\prime} (
\omega )}{(\hbar \omega)^{\eta}}
{}~~~~\mbox{at $T=0$}.
\label{defZ}
\end{equation}
The factor of 4 is for convenience in a later model calculation.\\
({\em ii\/}) and ({\em iii\/}) Two energy scales, $\Delta$ and $\Gamma$,
which measure the
pseudogap and
the strength of the damping respectively. These are obtained by solving the
constraints
\begin{equation}
\chi^{\prime \prime}_{L}
 ( \hbar \omega = \Delta ) = \frac{{\cal Z} M^{\eta}}{8}~~;~~
\frac{\Gamma}{\Delta^2} =
\frac{4 \pi}{{\cal Z} M^{\eta}} \lim_{\omega \rightarrow 0} \frac{
\chi_{L}^{\prime\prime}
(\omega )}{
\hbar \omega}~~~~\mbox{both at $T=0$}.
\label{defs}
\end{equation}
The factor of 8 in the first equation
is chosen to be exactly twice the factor 4 in (\ref{defZ}); the factor of
$4\pi$ in the
second equation is arbitrary and is for future convenience.
 The parameter $M$ is the energy below which the anomalous scaling of the
primary fixed point
$\chi_L^{\prime\prime} \sim \omega^{\eta}$ stops,
and a convenient choice is $M \equiv \alpha\mbox{max}(\Gamma, \Delta)$, where
$\alpha$ is of order, but larger than, unity. the factors of $M^{\eta}$
above also
ensure that $\Delta$ and $\Gamma$ have units of energy. Because of the tiny
value of
$\eta
\approx 0.03$, the
$M^{\eta}$ can be dropped while determining $\Gamma$ and
$\Delta$ from
data; this is however not true in any analytical computation of scaling
functions, as they are essential in
canceling cut-off dependencies and obtaining a universal result.
The pseudo-gap $\Delta$ has therefore been defined as the frequency at which
$\chi_L^{\prime\prime}$ falls to half its large frequency value (modulo
factors of $M^{\eta}$),
while
$\Gamma$ is determined from the sub-gap absorption.  For $\Gamma \leq
\Delta$, $\Delta$ is roughly the location of the
knee in $\chi_L^{\prime\prime} ( \omega )$, while $\Gamma$ is its width
(see Fig~\ref{fig1}b); however, the definitions above
hold for all $\Gamma / \Delta$.
For very small $\Gamma$, when the system is very close to the primary fixed
point,
$\Gamma$ and $\Delta$ measure the strength of the two relevant
perturbations away from it;
this is no longer true for large $\Gamma/\Delta$, but our scaling results
below will continue
to be valid.\\
({\em iv\/}) The final parameter sets the normalization of length scales. A
convenient choice is to
use the $T=0$ spin wave velocity, $c$, defined by
the $q$ dependence of the peak in the imaginary part of the
staggered spin
susceptibility
$\chi_{s} ( q,\omega )$ at large $q$ (In all expressions for $\chi_s$ {\em
only\/}, it is
implicitly assumed that wavevectors, $q$, are deviations from the ordering
wavevector
$Q$).

Scaling functions of the primary $z=1$ fixed point were studied in some
detail in Ref.~\cite{Ch-Sach},
and we can easily extend that analysis to write down the crossover scaling
functions for
the present case. The main difference is that we have a new relevant
parameter, $\Gamma$, which
must be included in the scaling functions.
We consider only case of the scaling function of the staggered
susceptibility, $\chi_s$, here,
leaving for later other observables which have a somewhat more complicated
crossover behavior.
Using the fact that $\Gamma$ was
defined above to have
the dimensions of energy, we can write
\begin{equation}
\chi_s ( q, \omega ) = \frac{{\cal Z}}{(k_B T)^{-\eta}} \left( \frac{\hbar
c}{k_B T} \right)^2 \Phi_s
\left( \overline{q}, \overline{\omega}, \overline{\Delta},
\overline{\Gamma} \right)
\label{cross}
\end{equation}
where $\Phi_s$ is a fully universal, dimensionless crossover function, and
the dimensionless arguments measure values
of the parameters in units of $k_B T$: thus
\begin{equation}
\overline{q} \equiv \frac{\hbar c q}{ k_B T}~~~;~~~
\overline{\omega} \equiv \frac{\hbar \omega}{ k_B T}~~~;~~~
\overline{\Delta} \equiv \frac{\Delta}{ k_B T}~~~;~~~
\overline{\Gamma} \equiv \frac{\Gamma}{ k_B T}.
\end{equation}
All letter exponents $\eta$, $\nu$ (to be used later) refer to the
primary fixed point.
The main condition for the validity of (\ref{cross}) is that the five
energy scales,
$\hbar c k$, $\hbar \omega$, $\Delta$, $\Gamma$ and $k_B T$ are all
significantly smaller than
upper cutoffs like the Fermi energy or an exchange constant. The ratios of
the five energy scales
can however be arbitrary.

At $\overline{\Gamma} = 0$, $\Phi_{s}$ reduces to the result of
Ref~\cite{Ch-Sach}
(there are some minor differences in conventions), and we obtain the
scaling function for case $A$.
The criticality of the secondary fixed point appears in the
$\overline{\Gamma} \rightarrow \infty$ limit.
Then $\Phi_s$ should collapse into a secondary scaling function
$\Phi_s^{S}$~\cite{crossover,nelson} with
the $\overline{\Gamma}$ argument removed. The other arguments, and the
overall scale, of $\Phi_s$
will be multiplied by powers of $\overline{\Gamma}$, so that $\Phi_s^{S}$
and its arguments have
scaling dimensions appropriate to the secondary fixed point.
To illustrate this point more clearly, let us assume that the structure of
the type $B$
fixed point is identical to the Gaussian $z=2$ fixed point obtained in
spin-density-wave mean-field
theory; then we will have
\begin{equation}
\lim_{\overline{\Gamma} \rightarrow \infty} \Phi_s ( \overline{k},
\overline{\omega}, \overline{\Delta}, \overline{\Gamma} ) =
\frac{1}{\overline{\Gamma}^{1-\eta}}
\Phi_s^{S} \left(
\frac{\overline{q}}{\overline{\Gamma}^{1/2}},  \overline{\omega},
\frac{\overline{\Delta}}{\overline{\Gamma}^{1/2}} \right)~~~\mbox{upto
logarithmic
corrections}.
\label{reduced}
\end{equation}
The logarithmic corrections arise because the $z=2$ critical point has
marginal perturbations in
$d=2$~\cite{hertz}. It is however important to note that these logarithmic
corrections, and indeed the entire leading critical behavior of the
secondary fixed
point,  are
contained completely within the primary universal scaling function $\Phi_s$.
Everything about the logarithmic terms
is universal and, in particular, there is no non-universal cutoff-dependent
argument to the
logarithms. This universality is a key point, and is a consequence of our
earlier choice to define the scaling
functions with respect to the primary fixed point.
In systems with $\Gamma$ so large that it is greater than a high energy cutoff
like the Fermi energy, this strong universality will not hold; however,
the large $\Gamma$
limit of $\Phi_s$ can still be used to describe the critical behavior, with the
understanding that the various numerical scale factors are not correct as
they are now
non-universal.

\section{Mean-field SDW calculations for the Hubbard model}
\label{sdw}

The remainder of the paper will make the rather abstract arguments of
Section~\ref{intro}
concrete by presenting explicit calculations in some simple models.
We start with the mean-field SDW theory of the disordering transition in doped
antiferromagnets. The simplest model which displays such transition is a
one-band Hubbard model given by
\begin{equation}
{\cal H} = -t\sum_{<i,j>} a^{\dagger}_{i,\sigma}a_{j\sigma} -
t^{\prime}\sum_{<i,j^{\prime}>} a^{\dagger}_{i,\sigma}a_{j^{\prime}\sigma} +
U \sum_i n_{\uparrow}n_{\downarrow} .
\label{hub}
\end{equation}
Here $j$ and $j^{\prime}$ label the nearest and the next-nearest neighbors,
respectively, and $n = c^{\dagger}c$ is the particle density.
Depending on the density of carriers and the ratio $t^{\prime}/t$,
the Fermi surface of free electrons can either be closed,
 in which case it is centered at $(0,0)$ and
located entirely inside the magnetic Brillouin zone, or it can be an open
Fermi surface which crosses the magnetic Brillouin zone boundary
(Fig.~\ref{fermi-surface}).
 We will not specify the values of $t$ and
$t^{\prime}$ for which the Fermi surface is open or closed for a particular
doping concentration,
but rather consider the critical behavior of the dynamic spin
susceptibility in both cases.

Let us  assume that the model has commensurate
antiferromagnetic long-range order
down to a transition point (this assumption may not hold for a particular
choice of $t$ and $t^{\prime}$, but
it should always hold  for
 related models with additional momentum dependence in $U$~\cite{si};
this momentum dependence however does not
lead to new physics near the transition, and we will not consider this
complication here).
 The $(\pi,\pi)$ ordering implies
that, e.g., the $z$ component of the spin-density operator, ${\vec S}(q) =
(1/2) \sum_k
a^{\dagger}_{k+q, \alpha}
{\vec \sigma}_{\alpha,\beta} a_{k,\beta}$, has a nonzero expectation value at
$q =Q \equiv (\pi,\pi)$. In the SDW approach~\cite{schrieffer},
 the relation $\langle \sum_k a^{\dagger}_{k+Q,
\uparrow}a_{k,\uparrow} \rangle =
-\langle \sum_k a^{\dagger}_{k+Q, \downarrow}a_{k,\downarrow}\rangle
 = \langle S_z \rangle$, is used to decouple the
 quartic term in (\ref{hub}).
 After decoupling and diagonalization of the quadratic form, one
obtains~\cite{chub-mus}
\begin{equation}
{\cal H_{SDW}} = {\sum_k}^{\prime} E^{c}_k c^{\dagger}_{k\sigma} c_{k\sigma} +
E^{d}_k d^{\dagger}_{k\sigma} d_{k\sigma}
\end{equation}
where primes on the summation signs indicate that it is over
the reduced magnetic Brillouin zone, and we introduced
\begin{equation}
E^{c}_k = E^{-}_k + \epsilon^{+}_k, ~E^{d}_k = - E^{-}_k +
\epsilon^{+}_k
\label{en}
\end{equation}
where $E^{-}_k = \sqrt{N_0^2 + (\epsilon^{-}_k)^2}$, and
$N_0 = U\langle S_z \rangle, ~\epsilon^{\pm}_k =
 (\epsilon_k \pm \epsilon_{k+Q})/2~,
\epsilon_k = -2t(\cos k_x + \cos k_y) - 4t^{\prime} \cos k_x \cos k_y$.
 We will refer to the quasiparticles described by $c$ and $d$
operators as conduction and valence fermions, respectively.
Finally, the self-consistency condition on $\langle S_z \rangle$
is
\begin{equation}
\frac{1}{U} =  {\sum_k}^{\prime} \frac{n^d_k - n^{c}_k}{E^{-}_k}.
\label{self_consist}
\end{equation}
Below we will assume that $t^{\prime}$ is negative. This is consistent
with numerical calculations~\cite{hybertsen,gooding2}
 and fits to the measured~\cite{arpes}
 shape of Fermi surface for
$YBa_2 Cu_3 O_{7}$.

We now describe the evolution of the Fermi surface with doping.
The shape of the Fermi surface is determined from $E^{c,d} = \mu$.
At half-filling, the system is an insulator ($|\mu| < N_0$),
all valence band states are occupied
and all conduction band states are empty.
At finite doping, the chemical potential moves into the
valence band. As the maximum of $E^{d}_k$ is located at $(\pi/2,\pi/2)$ and
symmetry related points, the Fermi surface first opens in the
 form of hole pockets located around these points (Fig~\ref{evol_fs}a).
 The subsequent evolution
of of the Fermi surface with doping in the SDW phase
depends on the ratio $t^{\prime}/t$. For
sufficiently large $|t^{\prime}|$,
the bottom of the conduction band also becomes smaller than
$\mu$ above some particular doping
concentration, and
electron pockets appear in the conduction band (Fig~\ref{evol_fs}b).
The energy
gap between the hole and electron pockets
shrinks gradually as $N_0$ decreases, and, at the transition point,
the electron and hole pockets merge at
 $(x_0,\pi - x_0)$
and symmetry related points, where $cos^2 x_0 = |\mu/4t^{\prime}|$, and the
value of $|\mu|$ is determined from the solution of
Eq.~(\ref{self_consist}) with
$N_0 \rightarrow 0$ (Fig~\ref{evol_fs}c).
It can be verified that this transition is of type $B$, and $Q$ can span
two points on
the Fermi surface in the paramagnetic phase. For small enough
$|t^{\prime}|$, the
situation is different: the size
of hole pockets continues to increase with doping,
and at some concentration $x_1$,
 the hole pockets extend to $(\pi, 0)$ and related points
while magnetic order is still present (Fig~\ref{evol_fs}d).
 At concentrations smaller than $x_1$, electron pockets
either do not appear at all, or they first appear, their size passes through a
maximum, and then they disappear at some doping concentration $x_2 < x_1$.
At some concentration $x_3 > x_2$,  the long-range
order finally disappears, but the Fermi
 surface at the transition is located inside magnetic Brillouin zone
boundary (Fig~\ref{evol_fs}e).
 The disordering transition  is
therefore of type $A$~\cite{Trugman}.

We now present results from a mean-field SDW calculation of the uniform and
staggered
susceptibilities; details can be found in the appendix. We shall mainly be
interested in the small $q$ and $\omega$ behavior of these responses when
$N_0$ is small. We shall find that an important feature of type $B$ systems is
that the limits $q, \omega \rightarrow 0$ and $N_0 \rightarrow 0$ do not
commute.
We consider the two cases separately.

\subsection{Type $A$}
The transverse, staggered, susceptibility is found to behave as
\begin{equation}
\chi_{s\perp} ( q, \omega ) = \frac{N_0^2}{\rho_s} \frac{1}{q^2 - \omega^2 /c^2
- i a \omega q N^{2}_0}
\label{chisperpA1}
\end{equation}
where $a,c$ are constants ({\em i.e.\/} finite as $N_0 \rightarrow 0$)
and $\rho_s \sim N_0^2 \sim (g_c - g)$ is a spin stiffness. It is clear that
the damping is negligible as $N_0 \rightarrow 0$, and that the
limits $q, \omega \rightarrow 0$ and $N_0 \rightarrow 0$ do
commute. The susceptibility is also that expected for a $z=1$ system.

In the quantum disordered phase we then
have $\chi_s \sim (q^2 - \omega^2 / c^2
+ \Delta^2 /c^2)^{-1}$, where the gap $\Delta$ vanishes at the transition
point.
The pole in $\chi_s$ corresponds to a positive energy, spin-1 particle-hole
bound state which has center of mass momentum near $Q$, and which lies below
the
bottom of the particle-hole continuum at these momenta.

The transverse, uniform susceptibility at $N_0 \rightarrow 0$ can be
represented as a sum of the interband and intraband contributions
\begin{equation}
\chi_{u \perp} = \chi_{u\perp}^{inter} + \chi_{u\perp}^{intra}
\label{chiuA}
\end{equation}
where we found that $\chi_{u\perp}^{inter} \sim O(N_0^2)$ and
$\chi_{u\perp}^{intra} \sim \text{const} + O(N_0^2)$.
So $\chi_{u\perp}$ is finite at the transition, as it should be when reaching
a Fermi liquid. However, the assertion by Millis~\cite{Millis} that the
spin-wave velocity should be given by a hydrodynamic expression~\cite{hydro}
$c_{sw}^2 = \rho_s / \chi_{u\perp}$
is seen to be incorrect:
$c_{sw} = c$ is in fact finite at the transition. The correct
relationship actually turns out to be
$c_{sw}^2 = \rho_s / \chi_{u\perp}^{inter} $.

\subsection{Type $B$}
The non-commutativity of the limits $q, \omega \rightarrow 0$ and $N_0
\rightarrow 0$ now leads to much more complex behavior than for type $A$.
Consider first, the transverse, staggered susceptibility at $q=0$; in
this case a relatively complete evaluation is possible, and we find
\begin{equation}
\chi_{s \perp} (0, \omega) \sim - \frac{N_0^2}{\rho_s} \,
\left[ \frac{\omega^2}{\sqrt{4N_0^2 - \omega^2}}
 + O(\omega^2) \right]^{-1}
\label{freq}
\end{equation}
where $\rho_s \sim N_0^2 \sim (g_c - g)$.
The first, singular term arises from interband processes and is due to the
 integration over  the region of
momentum space
where the Fermi surface crosses the magnetic Brillouin zone boundary; there is
no such point for type $A$ systems. The second term is the regular
contribution from the reminder of the momentum space, it is
similar to that found in type
$A$ systems. It is clear that the behavior of $\chi_{s\perp}$ is dramatically
different for $\omega \ll N_0$ and $\omega \gg N_0$.

The $q$ dependence of $\chi_{s\perp}$ only has a relatively
simple form in the limiting regions.
For $ q,\omega \ll N_0$
we have
\begin{equation}
\chi_{s\perp} (q,\omega)  = \frac{N_0^2}{\rho_s}~ \left[\left(q^2 -
\frac{\omega^2}{c^{2}}\right) - \frac{\omega}{N_0}
 (b_1 \omega - i b_2 \psi ({\vec q},\omega))\right]^{-1}
\label{chisperpB1}
\end{equation}
where $c$, $b_1$ and $b_2$ are constants, and $\psi ({\vec q},\omega) =
\psi (|q_x + q_y|, |q_x - q_y|,\omega)$ is a
symmetrical function of its first two
 arguments, which at $q \gg
\omega/t$
is {\it linear} in $q$. The explicit form of $\psi$ is presented
in the appendix.
We see that the
the lowest-energy excitations are overdamped even in the magnetically
ordered phase, although we indeed have a Goldstone mode at $q=0$.
Also, the renormalized spin-wave
velocity (defined in the limit of small $b_2$)
decreases as $c_{sw} \sim N_0^{1/2}$.
In the opposite limit, $ q, \omega \gg N_0$
(and for all $q, \omega$ at the transition point) both terms in
the last bracket in (\ref{chisperpB1}) become imaginary, and we find
\begin{equation}
\chi_{s\perp} (q,\omega) \sim (q^2 - \omega^2 /c^{2} - i \omega
\gamma)^{-1} .
\label{chisperpB2}
\end{equation}
 The magnon
excitations are now clearly overdamped.
In the SDW approximation we found that
$\gamma$ {\it does not} depend on the ratio $\omega/q$.
In the quantum disordered phase we then
have
\begin{equation}
\chi_s \sim (q^2 - \omega^2 / c^2  + i \omega \gamma + \Delta^2 /c^2)^{-1}
\label{chisperpB3}
\end{equation}
 where $\Delta$ is now a pseudo-gap and vanishes at the transition point.
Notice also that $c$ {\it is not} the velocity of the longest wavelength
spin-wave excitations in the ordered phase, though indeed it has the same
order.

The same form of susceptibility was proposed earlier by
 Barzykin {\em et al.}~\cite{pines} on phenomenological grounds.
  We will use Eqn.
(\ref{chisperpB3}) below to motivate the field-theory for the critical point
and
the paramagnetic phase.

As for type $A$, the transverse, uniform susceptibility is found to have two
contributions from interband and intraband processes
which now take the form
\begin{equation}
\chi_{u \perp} = \chi_{u\perp}^{inter} + \chi_{u\perp}^{intra}
\label{chiuB}
\end{equation}
where now $\chi_{u\perp}^{inter} \sim O(N_0)$ and
$\chi_{u\perp}^{intra} \sim \text{const} + O(N_0)$.
Again, $\chi_{u\perp}$ is finite at the transition, and the
relationship~\cite{Millis}
$c_{sw}^2 =\rho_s /
\chi_{u\perp}$ does not work as it would predict $c_{sw} \sim N_0$.
The correct relationship is
$c_{sw}^2 = \rho_s / \chi_{u\perp}^{inter}$.
In practice, however, it is difficult to find
conditions when one can neglect the $b_2$ term in (\ref{chisperpB1}).
Magnon excitations are then overdamped, and
the issue of the behavior of the spin-wave velocity near the transition
is not that relevant.
Notice also that  as shown in the appendix, the $O(N_0)$ contributions
in $\chi_{u\perp}^{inter}$ and $\chi_{u\perp}^{intra}$ cancel each other,
so that $\chi_{u\perp} = \text{const} + O(N^{2}_0)$.

\subsection{Relationship to results in the Shraiman-Siggia model}
\label{secss}
The main weakness of the above mean-field SDW computations
is that they underestimate the contribution of magnetic fluctuations, which
may destroy long-range order at a smaller doping.
Such fluctuations may be more completely accounted for by studying the phase
transition in the Shraiman-Siggia (SS) model~\cite{shraiman}, which
however has
other weaknesses to be described below. In this subsection we will review the
results of such a study~\cite{sss}, compare them with the above SDW
results, and make
some speculations on how they may be reconciled. Readers not interested in this
issue can skip this subsection without any loss of continuity.

The SS model is a continuum theory of interactions between mobile holes and
spin fluctuations in a $t$-$J$ model {\em i.e.\/} a model in which strong local
repulsion between electrons allows one to project out states with more than a
single-electron on a site. In the language of the Hubbard model, there is a
significant band-gap between the lower and upper Hubbard bands, and the SS
model
focuses exclusively on physics within the lower band. The lower and upper
Hubbard
bands can be identified with the valence and conduction bands,
respectively, of the
SDW theory discussed above. Furthermore, in the SDW mean-field theory, the
minimum,
direct  band-gap between the conduction and valence bands is proportional to
the
N\'{e}el order parameter $N_0$, and therefore vanishes when long-range order
disappears. This is, however, not the case in the SS model, in which $N_0$
and the
band-gap are independent parameters, and the band-gap is assumed to remain
finite
at the point where $N_0$ vanishes. This  is a key difference between the two
approaches.

Let us now approach the magnetic transition from the ordered side in the SS
model.
A simple computation, similar to those in Ref~\cite{sss}, shows that the
transverse,
staggered susceptibility has precisely the form (\ref{chisperpA1}) obtained
above
in the SDW theory of a type A transition; in particular the result
$\mbox{Im} (\chi_{s\perp})^{-1} \sim N_0^2 \omega q$ is also obtained in the
SS model (thus the proper interpretation of the factor of $N_0^2$ in
(\ref{chisperpA1}) in the SDW theory is that of the square of the magnetization
order parameter, rather than the valence/conduction band gap).
At the point where $N_0$ vanishes, computations in the SS model
show that~\cite{sss} $\chi_{s\perp} \sim (q^2 - \omega^2 / c^2 + i
a^{\prime} \omega^2 q )^{-1}$. In the mean-field SDW theory above we have
$\chi_{s\perp}
 \sim (q^2 - \omega^2 / c^2)^{-1}$ with no damping term; however it is quite
reasonable to expect that including higher-order paramagnon fluctuation
corrections in the SDW theory will lead to a damping term rather similar to
that
obtained in the SS model. So far, therefore, the spin-fluctuation
properties of the
magnetic transition in the SS model are essentially identical
to those obtained in the SDW theory of the $z=1$ transition in type A systems.

However, differences do appear when we consider the fermionic excitations.
The type A transition in the SDW theory has a large electron
Fermi surface wholly within the magnetic Brillouin zone, which changes
little between
the two phases on either side of the critical point. On the contrary,
 the Fermi surface of
the SS model, on the ordered side of the critical point,
 consists of elliptical hole pockets at the boundary of the
magnetic Brillouin
zone. The fate of
the elliptical
hole pockets in the quantum disordered phase of the SS model is not at all
clear~\cite{sss}.
Below we present a reasonable, but speculative, scenario:
we propose that in this phase, the
Fermi surface is
large ({\em i.e.} encloses a volume
equal to the total number of
electrons), as it was in both types of SDW transitions. However the
quasiparticle residue is very anisotropic so that at the transition to
N\'{e}el state,
 the residue vanishes
everywhere except for the regions which
surround the hole pockets in the ordered phase~\cite{varma}.
This implies that the critical
theory of the $z=1$ transition in the SS model in Ref~\cite{sss}
remains essentially correct.
There may
also be states in
the Hubbard gap  with their residue also
vanishing at the transition to N\'{e}el order.
If this large Fermi surface intersects
the magnetic Brillouin zone boundary
in the disordered phase, then there will be a
{\it finite\/} $T=0$ damping, $\gamma$, at $q=0$ in the quantum-disordered
phase; however
$\gamma$ will vanish faster than the gap, $\Delta$, as
one approaches the critical point, thus behaving as a dangerously irrelevant
perturbation (i.e., perturbation which is irrelevant very near transition but
leads to a new physics at some distance away from the transition).
 Alternatively, the large Fermi surface with anisotropic
quasiparticle
residue may be entirely within the magnetic Brillouin zone; in this case the
properties of the quantum disordered phase will be very similar to
those of a type $A$ transition in SDW theory.

\section{Model field theory for type B}
\label{largeN}

This section will present explicit computations of the scaling functions of
Sec~\ref{secscalinganalysis} in a model field theory appropriate for a
mean-field
type B transition.
The motivation for  the model follows the logic of
Hertz~\cite{hertz}: the action, $S$, for
paramagnon fluctuations in the quantum disordered phase should have
 a propagator which reproduces
the mean-field, type $B$, spin susceptibility  in (\ref{chisperpB2}).
The partition function of $S$ in
Matsubara imaginary time then has a form
\begin{eqnarray}
Z &=& \int {\cal D} \vec{n}(x, \tau) ~\delta (\vec{n}^2 (x,\tau)  - 1)
{}~\exp\left( - S[\vec{n}(x, \tau)] \right)
\nonumber \\
S = \frac{1}{2g} &T& \sum_{\omega_n} \int \frac{d^2 q}{4 \pi^2}
\left| \vec{n} ( q, \omega_n ) \right|^2 \left( q^2 + \omega_n^2 /c_0^2 +
\gamma |\omega_n |
\right) .
\label{action}
\end{eqnarray}
The action is written in Fourier space, and $\omega_n$ is a
Matsubara frequency. The vector field $\vec{n} (x, \tau)$ represents the local
orientation of the antiferromagnetic order parameter; we will allow
$\vec{n}$ to have
$N$ components to allow a subsequent large $N$ calculation. We have chosen
to implement
a fixed-length constraint $\vec{n}^2 =1 $ to mimic interactions between the
paramagnon
modes. This restriction is however not crucial and identical scaling
results would
be obtained in a model with a more conventional $(u/2N) (\vec{n}^2 )^2$
interaction.
The scaling
results appear a little more directly in the fixed-length model.
The first two terms in $S$ are those found
in the usual $O(N)$ non-linear sigma model. The last term
arises from the damping induced by the fermion particle-hole pairs.
The action $S$ was first explicitly written down by Liu
and Su~\cite{su}---they  used a two-component microscopic model
in which the spins and fermions are locally independent degrees of freedom,
and then integrated
the fermions out.

We expect $S$ to provide a reasonable description of the temperature
dependent crossovers in the
quantum disordered phase. However, close enough to the critical point in
the quantum disordered
phase, the validity of $S$ appears to us to not have been established
convincingly. It is likely
that it will be necessary to account for fluctuations involving incipient
formation of the
spin-density-wave gap over portions of the Fermi surface---note that these
gaps form at precisely
the same points on the Fermi surface which are responsible for the low
frequency
damping on the disordered side;
these issues will be examined in more detail in subsequent work. It is also
easy to see explicitly
that $S$ certainly breaks down on the
magnetically ordered side of the transition. On this side, the
spin fluctuation propagator changes substantially  when $q$ and $\omega$
become smaller than the order parameter $N_0$ (see e.g., Eq. (\ref{freq})),
 and these changes {\em cannot\/}
be implemented by simply introducing a condensate of the order parameter
into $S$.

In the remainder of Sec.~\ref{largeN} we will focus exclusively on the
model $S$. We will
describe its asymptotic critical behavior, and compute its scaling
functions in a large $N$ expansion.

It is also interesting to note in passing that the $|\omega|$ dissipation in
$S$ is similar to that used in the macroscopic quantum tunneling
literature~\cite{leggett}.
There the emphasis is on the consequence of this dissipation on the quantum
mechanics of a single, heavy particle, whereas we are studying its effects on
a field theory describing a large number of degrees of freedom.

Let us first discuss some general scaling properties of
$S$ from the vantage point of
the primary, $z=1$ fixed point. This fixed point clearly lies on the
submanifold of the
parameter space with $\gamma = 0$.
The term proportional to $\gamma$ is clearly a relevant
perturbation at the primary fixed
point. Moreover,  $|\omega_n |$ is non-analytic in frequency and leads to a
long-range $-1/\tau^2$
interaction in spacetime.
Theories with long-range interactions of this type are familiar in the
context of finite temperature
classical phase transitions~\cite{aharony} and many results can be
transferred over.
In particular, the $|\omega_n|$ term gets no singular loop
renormalizations, and the
scaling dimension
of $\gamma$ is exactly $1-\eta$. Therefore, at $T=0$, the response
functions will therefore be scaling functions of the ratio
$\gamma/(g-g_c )^{(1-\eta)\nu}$. The fully renormalized parameters
introduced earlier
will have a different dependence on the bare coupling constants depending
on the value of this ratio; this dependence can be deduced by standard
scaling arguments.
 For $\gamma \ll (g-g_c)^{(1-\eta)\nu}$ we will have dependencies
characteristic of the
primary fixed point for which
\begin{equation}
{\cal Z} \sim \mbox{constant},~~~~~ \Delta \sim (g-g_c)^{\nu},~~~~~
\Gamma \sim \gamma (g - g_c )^{\eta\nu}.
\end{equation}
In the opposite limit
$\gamma \gg
(g-g_c
)^{(1-\eta)\nu}$ we are  much closer to the critical point and dominated by
the secondary behavior; in
this case we have
\begin{equation}
{\cal Z} \sim \mbox{constant},~~~~~\Delta \sim
(g-g_c)^{1/2} \gamma^{(1-1/2\nu )/(1-\eta )},~~~~~\Gamma \sim
\gamma^{1/(1-\eta)}.
\end{equation}

Explicit results for
scaling functions of $S$ can be obtained in a $1/N$ expansion.
The methods are very similar to those
discussed at length in Ref.~\cite{Ch-Sach} for $\gamma = 0$; one only has to
add
a $\gamma |\omega_n |$ to each propagator. We will therefore merely present
the final
results.
At $N=\infty$ we found
\begin{equation}
\Phi_s^{N = \infty} ( \overline{q}, \overline{\omega} , \overline{\Delta},
\overline{\Gamma} )
= \frac{1}{\overline{q}^2 - \overline{\omega}^2 - i \overline{\Gamma}
\overline{\omega}
+ m^2 ( \overline{\Delta} , \overline{\Gamma} ) } .
\end{equation}
We chose the definition of the parameters earlier so that as $T \rightarrow
0$, $m =
\overline{\Delta}$. For finite $T$,  $m$ is given implicitly as a universal
function of $\overline{\Delta}$
and $\overline{\Gamma}$ by
\begin{equation}
\frac{\overline{\Gamma}}{2} \log \left( \frac{\overline{\Delta}^2}{m^2} \right)
+ \overline{\Gamma} \phi \left( \frac{m}{\overline{\Gamma}} \right)
- \overline{\Gamma} \phi \left( \frac{\overline{\Delta}}{\overline{\Gamma}}
\right)
 = 2 \int_0^{\infty} \frac{d \Omega}{e^{\Omega} - 1}
\tan^{-1} \left( \frac{\overline{\Gamma}
\Omega}{m^2 - \Omega^2} \right)
\label{meq}
\end{equation}
where the value of the arctangent runs from
$0$ to $\pi$ as $\Omega$ runs from $0$ to $\infty$, and
\begin{equation}
\phi(x) =  \left\{
\begin{array}{ll}
(4 x^2 - 1)^{1/2} \tan^{-1} (4 x^2 - 1)^{1/2} &~~~~\mbox{ for $x \geq
1/2$} \\
 & \\
{\displaystyle \frac{(1-4x^2)^{1/2}}{2} \log \frac{1 - (1 - 4x^2)^{1/2}}{1
+ (1- 4x^2
)^{1/2}} }&~~~~~\mbox{for
$x \leq 1/2$}
\end{array} \right. .
\label{defphi}
\end{equation}
Despite appearances to the contrary, $\phi (x) $ is analytic for all real $x$,
including $x=1/2$.
In the limit $\overline{\Gamma}
\rightarrow 0$, (\ref{meq}) reproduces the
$z=1$ result of Ref~\cite{Ch-Sach}: $m = 2 \sinh^{-1}
(e^{\overline{\Delta}/2}/2)$. In the opposite limit
$\overline{\Gamma} \rightarrow \infty$ we expect properties of the
secondary fixed point---we see from
(\ref{reduced}) that $m^2/\overline{\Gamma}$
should be a function of $\overline{\Delta}^2/\overline{\Gamma}$.
 This is exactly
what we find from (\ref{meq}) which
reduces to
\begin{equation}
\frac{m^2 - \overline{\Delta}^2}{\overline{\Gamma}} +
\frac{m^2}{\overline{\Gamma}}
\log \frac{\overline{\Gamma}^2}{m^2} -
\frac{\overline{\Delta}^2}{\overline{\Gamma}}
\log \frac{\overline{\Gamma}^2}{\overline{\Delta}^2} = 2 \int_0^{\infty} d
\Omega
\frac{\tan^{-1}(
\Omega /(m^2/\overline{\Gamma}))}{
e^{\Omega} - 1}
\label{millis}
\end{equation}
for large $\overline{\Gamma}$.
The only violation of the $\overline{\Delta}^2/\overline{\Gamma}$
scaling in the above comes from
the logarithms, which was also expected; the arguments of the logarithms
however remain universal.
Analysis of the solution of (\ref{millis}) for $m$ can be shown to yield
the same crossovers (including and upto $\log\log$ corrections) as those
discussed in Ref.~\cite{Millis} in the quantum disordered region for
$d=2$, $z=2$.

The calculations of the $1/N$ corrections also follow Ref~\cite{Ch-Sach}.
The evaluation requires substantial numerical computations which are currently
being carried out.
However, we also know from Ref~\cite{Ch-Sach} that at
least at $\overline{\Gamma} = 0$, these corrections were quite small, and
that the $N = \infty$ results were
satisfactory for most purposes
(a notable exception is $\chi^{\prime,\prime} (\omega)$
for $\Gamma \ll \Delta$, for which
 $1/N$ corrections are relevant).
The agreement of the large $\Gamma$,
$N=\infty$ results with those of Ref.~\cite{Millis}
is also encouraging.

A number of experimentally measurable quantities can be deduced from the
above results. For example,
the antiferromagnetic correlation length is
\begin{equation}
\xi = \frac{\hbar c}{m T},
\label{xieq}
\end{equation}
 and the
NMR relaxation rates (for a review, see, e.g., \cite{Slichter:review}),
upto known prefactors associated with hyperfine coupling
constants, are~\cite{pines,CSS}
\begin{equation}
\frac{1}{T_1} \propto \frac{\overline{\Gamma}}{m^2}~~~;~~~\frac{1}{T_{2G}}
\propto \frac{1}{ m T}.
\end{equation}
We plot in Fig~\ref{xifig} the universal crossover function for $\hbar c/ \xi
\Delta$ as a function of $k_B T/\Delta$ for
various values of $\Gamma / \Delta$. For $\Gamma \ll \Delta$ we have two
main regimes of temperature:
\begin{eqnarray}
&& (a)~ k_B T \gg \Delta~~~~~~~~~~\xi = 2
\log\left(\frac{\sqrt{5}+1}{2}\right)
\frac{\hbar c}{k_B T} \nonumber \\
&& (b)~ k_B T \ll \Delta~~~~~~~~~~\xi = \frac{\hbar c}{\Delta}~\left(1 - 2
\frac{k_B
T}{\Delta } \left(\exp\left(-\frac{\Delta }{k_B T}\right) + \frac{\pi}{6}
{}~\frac{k_B
T}{\Delta }~\frac{\Gamma}{\Delta}\right)\right).
\end{eqnarray}
Regime $(a)$ is the $z=1$ quantum-critical behavior~\cite{CHN,Ch-Sach},
and $(b)$ is the $z=1$ quantum-disordered.
At very low temperatures, the damping gives
rise
to a power-law rather than exponential behavior of the correlation length;
these power-laws are the same as for quantum disordered,
$z=2$ behavior, discussed
below. However, unlike for ``pure'' $z=2$ relaxational behavior,
$\chi''(\omega)$ has a knee at $\omega\sim \Delta\gg T$. Based on the
presence of such knee (shown in Fig.\ref{fig1}b),
we identify this regime as quantum disordered, $z=1$.

For $\Gamma \gg \Delta$ there are three sub-regimes:
\begin{eqnarray}
&& (a^{\prime})~ k_B T \gg \Gamma~~~~~~~~~~~~~~~~\xi = 2
\log\left(\frac{\sqrt{5}+1}{2}\right)
\frac{\hbar c}{k_B T} \nonumber \\
&& (b^{\prime})~ \frac{\Delta^2}{\Gamma} \ll k_B T \ll \Gamma~~~~~~~~\xi = f_1
\left(\frac{k_B
T}{\Gamma}\right) \frac{\hbar c}{(\Gamma k_B T)^{1/2}} \nonumber \\
&& (c^{\prime})~ k_B T \ll \frac{\Delta^2}{\Gamma}~~~~~~~~~~~~~~~~\xi =
\frac{\hbar
c}{\Delta}~ \left(1 - \frac{\pi^2}{12 \log(\Gamma/\Delta)}~\left(\frac{k_B
T}{\Delta }\right)^2~\left(\frac{\Gamma}{\Delta}\right)^2 \right).
\label{xires}
\end{eqnarray}
with $f_1 (x)$ a very slowly (logarithmically) varying, numerically
calculable function of
order unity.
Now regime $(a^{\prime})$ is $z=1$ quantum-critical, $(b^{\prime})$ is $z=2$
quantum-critical, and $(c^{\prime})$ is $z=2$ quantum-disordered.
In the negligibly small subregime of $(b^{\prime})$
where $\log\log(\Gamma / k_B T) \gg 1$, while maintaining $k_B T \gg
\Delta^2 / \Gamma$ we have
$\xi = \hbar c/(\pi \Gamma k_B T)^{1/2} [\log (\Gamma / k_B T)/
\log \log (\Gamma / k_B T)]^{1/2}$.
Most of the above asymptotic results are not terribly useful at any
realistic value
of $\Gamma / \Delta$, and the exact values plotted in Fig~\ref{xifig}
should be used for
experimental comparisons.

The above regimes are sketched in Fig~\ref{sketch}, which will be discussed
in Sec~\ref{conc}.
%
%
%
%
%
%

\subsection{Uniform susceptibility}
\label{secunisus}

Consider now the response of the nearly antiferromagnetic Fermi liquid to a
uniform magnetic field. Unlike the
staggered susceptibility which diverges at the transition, the uniform
susceptibility, $\chi_u$, is expected to remain
finite. As a result, the crossover behavior of $\chi_u$ is more complicated
and corrections to the leading scaling result now play a significant role.

The following results are for type $B$ systems; however they can also be
applied to type $A$ systems after imposing $\gamma=\Gamma=0$.

We need the modifications to the action $S$, in the presence
of an external field $H$. These corrections are equivalent to evaluating
the effective 3-point
and 4-point couplings between the antiferromagnetic order parameter and the
uniform
magnetization after integrating out the fermions; however, as we are mainly
interested in the
case of a uniform field, we will just carry the exact field dependence
through in all the
fermion propagators (see Sec~\ref{stagfield}).

In insulating antiferromagnets, the coupling to the field was
given exactly by a symmetry-related argument~\cite{conserve} with no
additional coupling
constants appearing. These symmetry arguments must now be applied with more
caution because of
the presence of gapless fermionic excitations in the quantum-disordered
phase. In particular, we
need to account for the continuous change in the Fermi liquid ground state
in the presence of
$H$ associated with the expanding (shrinking) of the up-(down-)spin Fermi
surface.

Consider, first, the effect of $H$ on the $|\omega_n|$ term in $S$. At small
$\omega_n$, this term arises from the absorption due to particle-hole pairs
with
momentum $Q$, made up of electrons and holes just above and below the Fermi
surface. A field,
$H$, will shift the bottom of the band of the up and down spins in opposite
directions.  In a model with a flat density of states, such a shift should
have {\em
no} effect on the particle-hole spectrum. So for small $H$, there is no
change in the
$|\omega_n|$ term. At larger $H$ there will be a change, because there is
always
some structure in the density of states: this will show up as a field
dependence
in the value of the coupling $\gamma$:
$\gamma \rightarrow \gamma ( 1 + \lambda \vec{H}^2)$ where $\lambda$ is a small
coupling. This is verified in an explicit computation of the field
dependence of $S$ discussed in the appendix.

The computation in the appendix also finds a precession term:
\begin{equation}
\frac{i}{g c_0^2} (1 + \alpha)
\int d^2 x d\tau \vec{H} \cdot \vec{n}
\times \frac{\partial \vec{n}}{\partial
\tau},
\label{field}
\end{equation}
where we have absorbed in the definition of $H$ a factor $g_G \mu_B /\hbar$
($g_G \mu_B$ is the gyromagnetic factor).
The prefactor of (\ref{field}) has been written such that the coupling
$\alpha=0$ in the insulating limit; in this limit the precession rate of
the spins
is known exactly and hence (\ref{field}) has no new coupling constants. In the
doped system, there is a correction to the precession rate from internal
fields generated by the polarization of the fermions, and this is represented
by
the new coupling $\alpha$. If the system is not too strongly doped, we can
expect
that $\alpha$ is not too large.

Finally, there is a term (also derived in the appendix),
 $\sim (\vec{n} \times \vec{H})^2$, which imposes an energetic
preference for the relative orientation of the antiferromagnetic order
parameter and the magnetic field.

All three terms discussed above induce non-universal
corrections in the $T=0$ value of
$\chi_u$. These may be interpreted as corrections to the fermionic Pauli
susceptibility from higher-order interactions between the paramagnon modes.

Things
do simplify, however, when we consider the $T$ dependent part of $\chi_u$.
Then the precession coupling in (\ref{field}) turns out to be the most
important in many cases.
The $T$-dependent corrections from the $H^2$ dependence of $\gamma$ can be
shown by
standard scaling arguments to be subdominant near the primary fixed point;
near the secondary fixed point this term turns out to have the same $T$
dependence
as that due to (\ref{field}), but with a prefactor of $(\gamma /t)^2$.
 Finally, the
last $(\vec{n}\times
\vec{H})^2$ term yields no temperature dependence in a model with a
fixed-length constraint. In a soft-spin theory, scaling arguments show that
this
term is also subdominant near the primary fixed point; near the secondary
fixed point, and going
beyond $N=\infty$ limit for soft-length spins, one obtains a $T$ dependence
similar
to that
due to the $H^2$ dependence of $\gamma$.

We see therefore, that if $\gamma$ is sufficiently large,
all three terms contribute roughly equally to the uniform susceptibility. The
temperature dependence of $\chi_u$ is then non-universal, and reliable
predictions for experiments are difficult.
To keep our discussion simple, and also because we think that this is most
realistic
experimentally, we will only present explicit results for $\chi_u$ for the case
when $\gamma$ is not too large. We will therefore only consider the $T$
dependence of
$\chi_u$ induced by the term
(\ref{field}) which dominates near the primary fixed point.

We now consider the consequence of (\ref{field}) on the $T$ dependence of
$\chi_u$.
Application of our earlier scaling arguments and the results of
Ref~\cite{Ch-Sach} allow us to deduce
the following crossover scaling function for $\chi_u$:
\begin{equation}
\chi_u (T) - \chi_u (T=0) = (1 + \alpha^{\prime} )\frac{k_B T}{c^2}
\Phi_{u} ( \overline{\Delta},
\overline{\Gamma} )
\label{alphaprime}
\end{equation}
where $\alpha^{\prime}$ is a non-universal constant, which
vanishes when $\alpha =0 $. The function
$\Phi_u$ is a completely universal scaling function which we will compute
below at $N=\infty$.
The only non-universalities of the above result are therefore related to
the coupling $\alpha^{\prime}$, and
the value of $\chi_u ( T=0 )$ which is dominated by the Pauli
susceptibility of the fermions.

The $N=\infty$ computation of $\Phi_u$ for $S$
 can be
performed as in Ref~\cite{Ch-Sach};
we find:
\begin{eqnarray}
\Phi_u^{N=\infty} (\overline{\Delta}, \overline{\Gamma} ) =
\frac{\overline{\Gamma}}{\pi^2} && \left [
\frac{1}{2} \log \left( \frac{\overline{\Delta}^2}{m^2} \right)
+ \frac{2 m^2 - \overline{\Gamma}^2}{4 m^2 - \overline{\Gamma}^2} \phi
\left( \frac{m}{\overline{\Gamma}} \right)
- \frac{2 \overline{\Delta}^2 - \overline{\Gamma}^2}{4 \overline{\Delta}^2
- \overline{\Gamma}^2} \phi \left(
\frac{\overline{\Delta}}{\overline{\Gamma}} \right) \right. \nonumber\\
 && ~~~~~~~~~~~~~~~~~~~~~~~~~~~~~~~~~~~\left. + 2 \int_0^{\infty} \frac{  d
\Omega}{e^{\Omega} - 1}
\frac{\Omega^3}{(m^2 - \Omega^2 )^2 + \overline{\Gamma}^2 \Omega^2 } \right]
\label{chieq}
\end{eqnarray}
where $m$ has to determined from (\ref{meq}) as a function of
$\overline{\Delta}$ and
$\overline{\Gamma}$,
and the function $\phi(x)$ was defined in (\ref{defphi}).
A plot of the universal contribution from $\Phi_u$ to $\chi_u$ is shown in
Fig~\ref{chifig}, with the
non-universal prefactor $1+\alpha^{\prime}$ dropped. There are several
different
regimes, similar to those found in the correlation length. Dropping the
$1+\alpha^{\prime}$, we state
the leading behavior of $ \chi_u (T)  $ in these regimes.
For $\Gamma \ll \Delta$ we have the two
regimes of temperature:
\begin{eqnarray}
&& (a)~ k_B T \gg \Delta~~~~~~~~~~\chi_u (T) - \chi_u (0) =
\frac{\sqrt{5}}{\pi}
\log\left(\frac{\sqrt{5}+1}{2}\right) \frac{k_B T}{c^2} \nonumber \\
&& (b)~ k_B T \ll \Delta~~~~~~~~~~\chi_u (T) - \chi_u (0) =
\frac{\Delta}{\pi c^2} ~\left(\exp\left(-\frac{\Delta }{k_B T}\right) +
\frac{\pi}{6}~\frac{\Gamma}{\Delta}\left(\frac{k_B T}{\Delta }\right)^2
\right).
\label{ex1}
\end{eqnarray}
For $\Gamma \gg \Delta$ the three sub-regimes are:
\begin{eqnarray}
&& (a^{\prime})~ k_B T \gg \Gamma~~~~~~~~~~~~~~~~\chi_u (T) - \chi_u (0) =
\frac{\sqrt{5}}{\pi}
\log\left(\frac{\sqrt{5}+1}{2}\right) \frac{k_B T}{c^2} \nonumber \\
&& (b^{\prime})~ \frac{\Delta^2}{\Gamma} \ll k_B T \ll \Gamma~~~~~~~~\chi_u
(T) -
\chi_u (0) = f_2 \left(\frac{k_B
T}{\Gamma}\right) \frac{k_B T}{c^2} \nonumber \\
&& (c^{\prime})~ k_B T \ll \frac{\Delta^2}{\Gamma}~~~~~~~~~~~~~~~~\chi_u (T) -
\chi_u (0) =
\frac{\Gamma}{6 \log ( \Gamma/ \Delta )}\left(\frac{k_B T}{\Delta c}\right)^2.
\label{ex2}
\end{eqnarray}
with $f_2$ a
function similar to the function $f_1$ above in (\ref{xires}). The regimes
$(a)$, $(b)$, $(a^{\prime})$, $(b^{\prime})$, $(c^{\prime})$ are the same
as those
described for the correlation length (notice that in the regime $(b)$, the
damping term gives rise to a power-law behavior at very low $T$).
Again, as with the correlation length, most of the the asymptotic
expressions are not useful
for realistic $\Gamma / \Delta$ and numerically determinable values of the
crossover
function should be used.
 Within the regime $(b^{\prime})$, it is
possible, as was the case with
$\xi$, to have  a negligibly small sub-regime where
$\log\log(\Gamma / k_B T)
\gg 1$ while $k_B T \gg \Delta^2 / \Gamma$; here we find $\chi_u (T) -
\chi_u (0) = (k_B T /
\pi c^2 ) (\log  \log(\Gamma / k_B T))/\log(\Gamma / k_B T)$.

\subsection{Specific Heat}
\label{secspecificheat}

The leading contribution to the free energy density of $S$ is strongly
divergent
in the ultraviolet $\sim \Lambda^3 $, where $\Lambda$ is an upper cutoff in
momentum space. In the model with $\gamma = 0$, it was found in
Ref.~\cite{Ch-Sach} that a single subtraction of the $T=0$ value of the
free energy
density was sufficient to make the $T$ dependent remainder finite. There were
no terms diverging with powers of $\Lambda$ smaller than 3---a direct
consequence of the hyperscaling properties of the original theory.
For finite $\gamma$, we find here that this simple property does not hold: a
single subtraction still leaves a term that is only logarithmically dependent
on
the upper cutoff. This is perhaps not surprising as the secondary fixed point
violates hyperscaling. While most of this violation
is actually cutoff by the crossover to the primary fixed point at high
energies,
some manages to survive.

We will restrict ourselves here to presenting the $T$ dependence of the free
energy density, ${\cal F}$, of the action $S$ in a $N=\infty$ calculation.
We found
\begin{equation}
{\cal F}(T) = {\cal F}(0)  + \frac{( k_B T)^3}{(\hbar c)^2} \Psi (
\overline{\Delta},
\overline{\Gamma} )
\end{equation}
where the dimensionless function $\Psi$, instead of being fully universal, has
a logarithmic dependence on $\Lambda/T$:
\begin{eqnarray}
\Psi (\overline{\Delta}, \overline{\Gamma} ) = &&
\frac{N}{4 \pi^2} \int_0^{\infty} \frac{  d
\Omega}{e^{\Omega} - 1}
\left[ (m^2 - \Omega^2) \tan^{-1} \left( \frac{\overline{\Gamma} \Omega}{
m^2 - \Omega^2} \right) + \frac{\overline{\Gamma}\Omega}{2}
\log \left(
\frac{(m^2 - \Omega^2 )^2 + \overline{\Gamma}^2 \Omega^2 }{(\hbar c\Lambda
/T)^4}
\right) \right]
\nonumber \\
&&~~~~~~~~~~~~~~ -\frac{N\overline{\Gamma}}{48 \pi^2} \left[
\left(3 m^2 - \frac{\overline{\Gamma}^2}{2} \right) \log
\frac{\overline{\Delta}^2}{m^2}
+ m^2 - \overline{\Delta}^2 + (4 m^2 - \overline{\Gamma}^2 )
\phi\left( \frac{m}{\overline{\Gamma}} \right) \right. \nonumber \\
&&~~~~~~~~~~~~~~~~~~~~~~~~~~~~~~~~~~~~~~~~~~~~~~~~~~~~~~\left. + (2
\overline{\Delta}^2 -6  m^2 + \overline{\Gamma}^2 ) \phi\left(
\frac{\overline{\Delta}}{\overline{\Gamma}} \right) \right]
\end{eqnarray}
where $m$ is determined from (\ref{meq}) as a function of $\overline{\Delta},
\overline{\Gamma}$ and the function $\phi$ is defined in (\ref{defphi}).

The asymptotic limits of the contribution of $\Psi$ to the specific
heat $C_V = - T \partial^2 {\cal F} / \partial T^2$ were
determined in a manner similar to $\xi$ and $\chi_u$.
For $\Gamma \ll \Delta$ we have:
\begin{eqnarray}
&& (a)~ k_B T \gg \Delta~~~~~~~~~~C_V = \frac{12 \zeta(3) N}{5 \pi}
k_B \left( \frac{k_B T}{\hbar c} \right)^2 \nonumber \\
&& (b)~ k_B T \ll \Delta~~~~~~~~~~C_V =
\frac{N \Gamma k_B ^2 T}{6 (\hbar c)^2}  \log \frac{\hbar c\Lambda}{\Delta}.
\end{eqnarray}
For $\Gamma \gg \Delta$ the three sub-regimes are:
\begin{eqnarray}
&& (a^{\prime})~ k_B T \gg \Gamma~~~~~~~~~~~~~~~~C_V = \frac{12 \zeta(3)
N}{5 \pi}
k_B \left( \frac{k_B T}{\hbar c} \right)^2 \nonumber \\
&& (b^{\prime})~ \frac{\Delta^2}{\Gamma} \ll k_B T \ll \Gamma~~~~~~~~C_V = N
f_3
\left(\frac{k_B T}{\Gamma}\right) \frac{\Gamma k_B^2 T}{(\hbar c)^2}
\nonumber \\
&& (c^{\prime})~ k_B T \ll \frac{\Delta^2}{\Gamma}~~~~~~~~~~~~~~~~C_V =
\frac{N \Gamma k_B ^2 T}{6 (\hbar c)^2} \log \frac{\hbar c \Lambda}{\Delta}.
\end{eqnarray}
with $f_3$ a
function similar to the function $f_1$ above in (\ref{xires}). The regimes
$(a)$, $(b)$, $(a^{\prime})$, $(b^{\prime})$, $(c^{\prime})$ are the same
as those
described for the correlation length.
Within the regime $(b^{\prime})$
above, it is
possible, as was the case with
$\xi$, to have  a negligibly small sub-regime where
$\log\log(\Gamma / k_B T)
\gg 1$ while $k_B T \gg \Delta^2 / \Gamma$; here we find $C_V = (N/12)
(\Gamma k_B^2 T/(\hbar c)^2) \log((\hbar c \Lambda)^2 / k_B T \Gamma)$.

\section{Conclusions}
\label{conc}

We begin this concluding section with
a simple, qualitative discussion of the physical picture behind the
computations in this and related, previous, works on the spin fluctuation
properties of the not too strongly doped cuprate compounds. A discussion of
the relationship of our ideas to experimental systems will follow.
Finally we will discuss some open theoretical issues.

It is very useful to think
in terms of the physics at different
length and energy scales. On the whole, we may assume that the relevant energy
scales decrease uniformly with increasing length scales, so the two can be used
interchangeably to move between the different regimes.
In what follows, we describe the sequence of crossovers as one moves
from larger to smaller energy scales, or equivalently from smaller to larger
length scales.

At the very largest
energy scales, the behavior is
 dominated by lattice scale physics, which is inherently non-universal. At
slightly smaller energies, provided the doping is not too large, one can
neglect the effect of mobile carriers. It was argued in Ref~\cite{Ch-Sach}
that spin
fluctuations at these scales are {\em ``quantum-critical'' \/} and well
described by
properties
of the $z=1$ critical point in the ${\rm O}(3)$ non-linear sigma model which
separates
the N\'{e}el ordered and magnetically disordered phases. This proposal has
been the source
of some controversy in the literature, although some fairly convincing
evidence has appeared
in recent
high temperature
series studies~\cite{rajiv}.
The excitations in
this $z=1$ quantum-critical regime are neither spin waves of the ordered
state, nor
$S=1$ magnons of the disordered state, but form a critical continuum.

The subsequent crossovers at smaller energies
depend on the doping concentration.
At very low doping, the presence of $T=0$ long-range order
in the Heisenberg  antiferromagnet becomes apparent; there is then a crossover
from $z=1$ critical
fluctuations at larger energies to the Goldstone spin-wave modes of the
ordered state at lower energies.
At a slightly larger doping, the $T=0$ long-range order is destroyed.
The
crossover at smaller energy scales
is then into the quantum-disordered state of the sigma model, where
the
excitations are then gapped, triply degenerate, spin-1 magnons. At this same
doping, and provided the geometry of the Fermi surface is of type $B$,
there is a
last crossover, at an even smaller energies, to a state
where the mobile carriers
induce an important $T=0$ damping of the magnon
excitations~\cite{hertz,Millis};
this damping leads to to sub-gap absorption and power-law
(in $T$) behavior of
observables. All power-laws are equivalent to those
of quantum disordered, $z=2$ regime, discussed below.

This sequence of crossovers has been  described for the case
in which the
damping is not too large: $\Gamma < \Delta$ in
 the notation of Sec~\ref{secscalinganalysis}.

At even larger dopings we may have $\Gamma \geq \Delta$; then
the intermediate quantum disordered, $z=1$, regime of a pure
sigma-model is not realized, and the system undergoes crossovers from $z=1$
critical regime to $z=2$ critical regime, and then at even smaller energies
to $z=2$ quantum disordered regime with relaxational behavior of spin
excitations.

The scaling regimes and associated crossovers are sketched in
Fig~\ref{sketch}. We emphasize that all phase boundaries between phases
on this sketch are smooth crossovers, and their
precise positions are therefore not meaningful, while the sequence
of crossovers and overall structure of the diagram
is meaningful.
In terms of this phase diagram, the Shraiman-Siggia model (type $A$),
corresponds to $\Gamma/\Delta=0$~\protect\cite{Ch-Sach,sss} (modulo
dangerously irrelevant
damping terms);  models with primarily relaxational
behavior of spin excitations (type $B$),
similar to that observed in $YBa_2Cu_3O_7$,
to $\Gamma/\Delta\agt 1$
\protect\cite{MMP,Millis}; and, finally, models with
moderate $\Gamma/Delta\alt 1$ exhibit spin
pseudogap behavior similar to that observed in the
underdoped Y-based materials,
such as $YBa_2Cu_3O_{6.63}$ and $YBa_2Cu_4O_8$ \protect\cite{Sokol,pines}.
The possible relationship of the phase boundaries to doping levels in
the cuprate
materials will be discussed in more detail in a subsequent publication.

Going to a finite $T$ introduces an additional degree of complication which
substantially changes the crossovers described above.
However, all of these $T$
dependent crossovers are contained in the universal crossover function
$\Phi_s$ in Eqn
(\ref{cross}), which was computed in a large $N$ expansion in this paper.
The main
results can be understood by keeping a simple rule of thumb in mind: the
primary effect
of a finite
$T$ is to thermally quench the excitations at the energy scale
$k_B T$, so that
the crossovers at scales below $k_B T$ no longer occur.
It is therefore
possible to make experimental predictions for temperature dependences
that are remarkably
universal~\cite{jinwu,Ch-Sach,Millis}.

We now briefly describe the relationship of our results to experiments on
high-$T_c$ oxides. We leave detailed quantitative comparisons
for a separate publication
which will also contain computations of $1/N$ corrections;
here we underline only the main qualitative points.

{\em $La_{2-x}Sr_xCuO_4$:}
The Fermi-surface of $La$-based
materials has not been measured experimentally. Apparently,
they are close to type A~\cite{si} in the terminology of
Sec~\ref{intro}, and we expect to see only
the features of the $z=1$ behavior.
However, these materials also show
nontrivial spin
correlations in the metallic phase
which will certainly effect the spin fluctuations at low
enough temperatures~\cite{aeppli}.  The primary effect of mobile
carriers on the uniform susceptibility should be a temperature-independent
additive contribution, which increases with doping.
We have suggested earlier~\cite{Ch-Sach,CSS,CS}
that this behavior may have been observed at a few percent $Sr$
concentration.

{\em $YBa_2Cu_3O_{7}$:}
Photoemission experiments~\cite{arpes} indicate that the
Fermi surface is large and  belongs to type $B$.
Both our and earlier \cite{MMP}
analysis of the NMR data is roughly consistent with type $B$ scaling
behavior at $\Delta/\Gamma\sim 1$:  the measured
$1/T_1 T$ monotonically decreases with $T$, and the ratio
$T_1 T/T^{2}_2$ is temperature independent \cite{Imai7}.
A quantitative
comparison with the measured uniform susceptibility
\cite{Johnston}
is difficult because
nonuniversal contributions to susceptibility,
neglected in (\ref{ex1}), (\ref{ex2}), are relevant for this
 fully doped material.

{\em $YBa_2Cu_3O_{6.6}$ and $YBa_2Cu_4O_{8}$:} For both of these materials,
the measured
Fermi surface is large and  belongs to type $B$~\cite{tranquada,photo}.
The measured $1/T_1$~\cite{Takigawa6,124T1} and uniform susceptibility
{}~\cite{Johnston} decrease rapidly as temperature decreases below
$T\simeq 150K$.
Such behavior is characteristic of
the quantum-disordered $z=1$ regime and we therefore expect that
for these materials, $\Gamma/\Delta\ll 1$.
Note however, that we have shown explicitly that due to the finite damping,
 this decrease, particularly for the uniform susceptibility,
 is not purely exponential as it is in phenomenological ``spin-gap'' models.
We also note that it is important to include $1/N$
corrections~\cite{Ch-Sach} in describing the
behavior of $1/T_1 T$ in this regime.
At high temperatures, this system is expected to crossover to $z=1$
quantum-critical
behavior, where $T_1T$ is linear in $T$ {\em and} $T_1T/T_2=\text{const}$
\cite{Ch-Sach,Sokol}; this behavior appears to have been
observed in both $YBa_2Cu_3O_{6.6}$ and $YBa_2Cu_4O_{8}$
\cite{Takigawa6,124T2}.

We conclude this paper by recalling and raising some open theoretical
questions.
The explicit computation of temperature-dependent crossovers has been
restricted to the quantum
disordered phase, and not too close to the quantum-critical point.
We suggested that $S$ may fail close enough to the critical point, and
saw explicitly that it failed badly in describing the low
energy
excitations on the ordered side. A drastic change in the action seems to be
necessary; probably,
much more drastic than adding some additional dangerously irrelevant
coupling which
becomes important only the ordered side.
In fact, any procedure which uses a truncated power series expansion in
$\vec{n}$ for
the action appears doomed: no such expansion will correctly capture the
appearance of
gaps on portions of the Fermi surface. It seems that the only natural way of
describing the crossover on the ordered side is to work in a theory in
which the fermions
are not completely integrated out. It is then possible to obtain an action
which has only regular terms, and which holds on both sides of the
transition. A possible form of the action involves two
species of fermions (corresponding to pairs of points on the Fermi surface
separated by $Q$)
coupled
to the $n$-field. Computing crossovers for such an action remains an
important open problem.

A complementary open problem applies to the analysis using the
Shraiman-Siggia~\cite{sss} model. In this case the ordered state and its
crossovers are
reasonably well studied. However it is also clear that the model cannot
apply in
detail on the quantum disordered side, and the crossovers between the
ordered and
disordered sides are not completely understood.

Finally, we make a few remarks about the effects of disorder.
Hertz~\cite{hertz} did
begin to address the consequences of disorder, but his analysis is
seriously incomplete
in all cases. One sign of this is that his value of $\nu$ (which equals 1/2
in all
the cases considered by him) violates the inequality $\nu >
2/d$~\cite{harris} where $d <
4$ is the spatial dimension. It is easy to determine the source of this
violation:
the most important perturbation on a soft-spin version of an action like
$S$ is a
random-``mass'' term $\int d^d x d\tau \sim m^2 (x) \vec{n}^2 (x, \tau)$
which accounts
for fluctuations in the local value of the critical coupling at which the
magnetic order is destroyed. It can be easily shown that this term is
relevant about the pure fixed point as long as $\nu < 2/d$. A
study of $S$,
with an additional random mass, along the lines of the work of Boyanovsky and
Cardy~\cite{cardy} should be quite straightforward. However this approach
involves the potentially
dangerous procedure of expanding about a problem with $\epsilon_{\tau}$
time dimensions,
and a method which works directly with $\epsilon_{\tau} = 1$ would be
preferable.

\paragraph*{Note added:} Just before this paper was submitted, we received a
preprint from L. B. Ioffe and A. J. Millis~\cite{IM} which addressed the $T$
dependence of the uniform susceptibility near the type B fixed point.
Their expression for $\chi_u$ in terms of correlators of
$\vec{n}$ is in general
agreement with our discussion in Sec~\ref{secunisus}. However, they
focused on the
temperature dependence of $\chi_u$ at very low $T$ ($\log \Gamma/T \gg 1$)
  near the type B fixed point;
we chose to focus on the term important near the primary type A fixed
point---the latter term
gives a universal (up to an overall
prefactor) contribution to $\chi_u$ which should be dominant at all $T$
unless
$\Gamma/\Delta$ is very large.

\acknowledgements
We thank E. Dagotto, L. B. Ioffe, A. J. Millis, and S. Trugman for helpful
discussions.
This research was supported by NSF Grant No.  DMR92-24290 (S.S.),
and the Graduate School at UW-Madison (A.C.).

\appendix
\section*{Susceptibilities in SDW theory at T=0}
\label{appenSDW}

\subsection{Staggered Susceptibility}

In the SDW theory, the spin susceptibility is given by a ladder series
of bubble diagrams.
At $T=0$, one fermion in the bubble should be above the Fermi surface, and one
below. In principle, the solution of the
ladder series in the magnetically ordered
state is a $2\times2$ matrix problem, as one has to consider bubbles with
momentum
transfer $0$ and $Q$~\cite{schrieffer,chub-fren}.
 For our considerations, it turns out
that all terms with momentum transfer $Q$
disappear at the critical point, and we have checked that they only account for
small corrections to the expressions obtained below. We therefore neglect
terms $\chi^{+-}_0 (q,q+Q,\omega)$ with the momentum transfer $Q$
 in which case the total transverse
dynamic susceptibility is given by
\begin{equation}
\chi^{+-} (q,\omega) = \frac{\chi_0^{+-} (q,\omega)}{ 1 - U\chi_0^{+-}
(q,\omega)}
\label{chitot}
\end{equation}
where
\begin{eqnarray}
\chi^{+-}_0 (q,\omega) &=& \frac{1}{2N} {\sum_k}^{\prime} \left[1 -
\frac{\epsilon^{-}_k \epsilon^{-}_{k+q} - N_0^2}{E^{-}_k
E^{-}_{k+q}}\right] \left(\frac{n^{d}_{k+q} - n^{c}_k}{E^{c}_k - E^{d}_{k+q} -
\omega} +
\frac{n^{d}_{k} - n^{c}_{k+q}}{E^{c}_{k+q} -E^{d}_k +  \omega}\right) +
\nonumber \\
&&  \frac{1}{2N}
{\sum_k}^{\prime}
 \left[1 +
\frac{\epsilon^{-}_k \epsilon^{-}_{k+q} - N_0^2}{E^{-}_k
E^{-}_{k+q}}\right] \left(\frac{n^{d}_{k+q} - n^{d}_k}{E^{d}_{k} - E^{d}_{k+q}
-\omega} + \frac{n^{c}_{k} - n^{c}_{k+q}}{E^{c}_{k+q} - E^{c}_{k} +
\omega}\right) .
\label{chi0}
\end{eqnarray}
Here and below we set $\hbar =1$. We will now study the behavior of this result
near the magnetic transition where $N_0$ becomes small.

Near $q =Q$, the coherence factors can be simplified to
\begin{eqnarray}
1 - \frac{ \epsilon^{-}_k \epsilon^{-}_{k+q} - N_0^2}{E^{-}_k
E^{-}_{k+q}} &\approx& 2 - (\epsilon^{-}_k + \epsilon^{-}_{k+q})^2
\frac{N_0^2}{2 (E^{-}_k)^4}  = 2 - O((Q-q)^2) \nonumber \\
1 + \frac{ \epsilon^{-}_k \epsilon^{-}_{k+q} - N_0^2}{E^{-}_k
E^{-}_{k+q}} &\approx& (\epsilon^{-}_k + \epsilon^{-}_{k+q})^2
\frac{N_0^2}{2 (E^{-}_k)^4} = O((Q-q)^2) .
\end{eqnarray}
We first consider the behavior of dynamic susceptibility at the
antiferromagnetic
momentum, $q = Q$. Here
 $\epsilon^{-}_k + \epsilon^{-}_{k+q} =0,~E^{d}_k + E^{c}_{k+q} =
2 E^{-}_k$, and one obtains using the self-consistency condition
\begin{equation}
\chi^{+-}_0 (Q,\omega) = \frac{1}{U} + \frac{1}{N} {\sum_k}^{\prime}
(n^{d}_{k} - n^{c}_k) \left(\frac{1}{2E^{-}_k -\omega} +
\frac{1}{2E^{-}_k + \omega} - \frac{1}{E^{-}_k}\right) .
\end{equation}
Expanding this expression in $\omega$ and substituting into (\ref{chitot})
we obtain
\begin{equation}
(\chi^{+-} (Q,\omega))^{-1} = \frac{U^{2}\omega^2}{4 N} {\sum_k}^{\prime}
\frac{n^{d}_k - n^{c}_k}{(E^{-}_k)^3} .
\label{freq_exp}
\end{equation}
The difference between type $A$ and $B$ transitions now becomes transparent.
 In the
first case, the integration over $k$ is restricted to a region which is
located entirely
 inside the magnetic Brillouin zone. Accordingly, $E^{-}_k$ remains finite
when $N_0$ tends to zero, and  at the critical point we have
$(\chi^{+-} (Q,\omega))^{-1} \sim \omega^2$. For type $B$, however,
the allowed region of
momentum integration includes the vicinity of
 $(x_0, \pi - x_0)$ and symmetry
related points where the Fermi surface crosses the magnetic Brillouin zone
boundary.
 At each of these points, $E^{-}_k = N_0$, i.e, the denominator in
(\ref{freq_exp})
diverges as $N_0 \rightarrow 0$. Expanding $E^{-}_k$ near these points
and performing the momentum integration, we obtain after some algebra
\begin{equation}
(\chi^{+-} (Q,\omega))^{-1} = - D \frac{U^2}{t^2}
 \left( \frac{\omega^2}{\sqrt{4N_0^2 - \omega^2}}
 + \frac{\omega^2}{c^2} \right)
\label{chi_pi}
\end{equation}
where $D = t/8 \pi |t^{\prime} \sin x_0 \sin
2x_0|$, and the $\omega^2$
 term comes from the integration over the regions far from
$(x_0, \pi - x_0)$.
We see that the leading term in the expansion in $\omega$ is now $\omega^2
/N_0$.
As one approaches the critical point, the first term in
(\ref{chi_pi}) becomes purely imaginary,
 and at $N_0 =0$ one has
\begin{equation}
(\chi^{+-} (Q,\omega))^{-1} = - D \frac{U^2}{t^2}
 \left( i \omega +
\frac{\omega^2}{c^2} \right) .
\label{exp_omega}
\end{equation}

We now consider the form of the susceptibility at finite $q$ which we will
understand as a deviation from $Q$.
The calculations here are straightforward but tedious, so we skip
some of the details.
First, we found that at $\omega=0$, all potential terms of
the form ${q}^2/N_0$ cancel each other
for both type $A$ and $B-$type transitions. The
expansion of $(\chi^{+-} (q, 0))^{-1}$ over ${q}$
is therefore regular for both cases. Second, we found that
when both $\omega $ and ${q}$ are finite, the second
piece in (\ref{chi0}) has an imaginary part which for
type $A$ transition behaves as
\begin{equation}
\mbox{Im} (\chi^{+-})^{-1} \sim \omega {q} N^{2}_0 .
\end{equation}
We see that the damping of magnetic
excitations disappears as $N_0 \rightarrow 0$.

For type $B$ transitions, imaginary part of the susceptibility is
a function of ${ q}/N_0$ and $\omega/N_0$. For
$q,\omega \ll N_0$, we obtained
\begin{equation}
\mbox{Im} (\chi^{+-})^{-1} = {\bar D}~\frac{U^2}{t N_0}
\psi(|q_x + q_y|, |q_x - q_y|,\omega)
\label{a}
\end{equation}
where  ${\bar D} = t/[8\pi|t^{\prime} \sin 2x_0|]$, and $\psi$
 looks particularly
simple at $t^{\prime} \ll t$
\begin{equation}
\psi (|q_x + q_y|, |q_x - q_y|,\omega) =
\frac{1}{2} ~\mbox{Re}~\left( \frac{(q_x + q_y)^4}{((q_x + q_y)^2 - {\bar
\omega}^2)^{3/2}} + \frac{(q_x - q_y)^4}{((q_x - q_y)^2 - {\bar
\omega}^2)^{3/2}}\right)
\end{equation}
where $\bar \omega = \omega/(2t \sin x_0)$. We also obtained expression for
$\psi$ for arbitrary $t^{\prime}/t$, but it is more cumbersome and we refrain
from presenting it here.
 In the opposite limit,
$q,\omega \gg N_0$, or equivalently when
$N_0$ vanishes, the ${ q}/N_0$
dependence in
(\ref{a}) transforms into a constant term, and  both
pieces in (\ref{chi0}) contribute linear in $\omega$ terms into the imaginary
part of $\chi^{+-}$. The numerical factors for each of the two pieces
contain combinations of $\theta-$ functions of the form, e.g.,
 $\theta (4t \sin x_0 |q_x + q_y| - |\omega + 4 |t^{\prime} \sin 2 x_0| (q_x -
q_y)|)$.
 However, we have checked explicitly
that the sum of the two terms does not contain $\theta$ functions.
Specifically, we obtained at $N_0 =0$ and {\it arbitrary} ratio $\omega/q$
\begin{equation}
\mbox{Im} (\chi^{+-})^{-1} = {\bar D}~\frac{U^2}{t^2} \omega .
\label{b}
\end{equation}

Combining the above results, one obtains the expressions presented in Eqns
(\ref{chisperpA1}), (\ref{chisperpB1}), (\ref{chisperpB2}).

\subsection{Uniform susceptibility}

The uniform transverse spin susceptibility is defined as
\begin{equation}
\chi_{\perp} = (1/2)(g_G \mu_B /a)^2 \lim_{q \rightarrow 0}
\chi^{+-} (q,\omega =0),
\end{equation}
 where $g_G \mu_B$ is the
gyromagnetic factor and $a^2$ is the volume per spin.
In the $q \rightarrow 0$ limit, the coherence
factors are simplified and we obtain $\chi_{\perp} = (g_G \mu_B /a)^2
\chi_0/[2(1 - U \chi_0)]$ where
\begin{equation}
\chi_{0} =  \frac{1}{N} {\sum_k}^{\prime} \left(\frac{N_0}{E^{-}_k}\right)^2
\left(\frac{n^{d}_{k} - n^{c}_k}{E^{-}_k}\right)
+  \lim_{q \rightarrow 0} \frac{1}{N}
{\sum_k}^{\prime}
\left(\frac{\epsilon^{-}_k}{E^{-}_k}\right)^2
\left(\frac{n^{d}_{k+q} - n^{d}_k}{E^{d}_{k} - E^{d}_{k+q}} +
 \frac{n^{c}_{k} - n^{c}_{k+q}}{E^{c}_{k+q} - E^{c}_{k}}\right) .
\label{chi_perp}
\end{equation}
As one approaches the critical point, the first term in (\ref{chi_perp})
disappears and $\chi_{\perp}$ takes the familiar RPA form of magnetic
susceptibility in a Fermi-liquid, which is finite at $T=0$.
For the type $A$ transitions, the first term vanishes as  $O(N_0^2)$, and
for  the type $B$ transitions it vanishes as $O(N_0)$.
The uniform susceptibility then has forms presented in Eqs. (\ref{chiuA}) and
(\ref{chiuB}). Notice however, that the full $\chi_0$ (and hence,
$\chi_{\perp}$) behaves as $\chi_0 = \text{const} + O(N^{2}_0)$ for both types
of transitions: for the type $B$ transition, linear in $N_0$ contributions from
the first and second terms in (\ref{chi_perp}) cancel each other.

\subsection{Staggered susceptibility in a finite magnetic field}
\label{stagfield}

To obtain the temperature dependent correction to the uniform susceptibility
from the fluctuation and interaction of the magnon modes, we will need the
form of the
long-wavelength action for antiferromagnetic fluctuations in the presence
of the external
magnetic field. To derive this form, we need
to understand how the staggered susceptibility is modified in such a field.
 In this subsection, we will determine this modification in the
SDW formalism for the ordered state. We then extend SDW calculations
upto a critical point which in turn will allow us to obtain the field-dependent
part of the action in a disordered phase.

To be definite, we assume that the field is
 directed along $x$ axis (antiferromagnetic order parameter is
along $z$ axis).  We also introduce new operators
as linear combinations of quasiparticle operators with up and down spins
\begin{equation}
\phi_k = \frac{a_{k,1} + a_{k,2}}{\sqrt 2};~~ \psi_k =  \frac{a_{k,1} -
 a_{k,2}}{\sqrt 2}
\end{equation}
(indices 1 and 2 correspond to up and down spin, respectively).
The quadratic part of the Hubbard Hamiltonian now takes a form
\begin{eqnarray}
{\cal H_2} &=&  {\sum_{k}}^{\prime} (\epsilon_k - {\bar H})
\phi^{\dagger}_k \phi_k +
(\epsilon_{k+Q} +  {\bar H})\psi^{\dagger}_{k+Q} \psi_{k+Q} - N_0
(\phi^{\dagger}_k \psi_{k+Q} + \psi^{\dagger}_{k+Q} \phi_{k}) + \nonumber \\
&& {\sum_{k}}^{\prime} (\epsilon_{k+Q} -  {\bar H})
\phi^{\dagger}_{k+Q} \phi_{k+Q}
 + (\epsilon_{k} +  {\bar H})\psi^{\dagger}_{k} \psi_{k} - N_0
(\phi^{\dagger}_{k+Q} \psi_{k} + \psi^{\dagger}_{k} \phi_{k+Q})
\label{quadr_form}
\end{eqnarray}
where $ {\bar H} =  H/2$.
Clearly,  the magnetic field splits
conduction and valence bands for up and
down spins so we now have four different branches of fermionic excitations.
However, we see that the pairs of operators $(\phi_k, \psi_{k+q})$ and
$(\phi_{k+Q}, \psi_k)$ in (\ref{quadr_form}) are decoupled from each other,
and therefore
the diagonalization is still a $2\times 2$ problem. Performing the
standard manipulations we obtain
\begin{equation}
{\cal H_{SDW}} = {\sum_k}^{\prime} E^{\alpha}_k \alpha^{\dagger}_{k}
\alpha_{k} +
E^{\beta}_k \beta^{\dagger}_{k} \beta_{k} + E^{\gamma}_k
\gamma^{\dagger}_{k} \gamma_{k} +
E^{\delta}_k \delta^{\dagger}_{k} \delta_{k}
\end{equation}
where we introduced
\begin{equation}
E^{\alpha,\beta} = \epsilon^{+}_k \pm E^{--}_k, ~E^{\gamma,\delta}
= \epsilon^{+}_k \pm E^{-+}_k
\label{enh}
\end{equation}
and $E^{--}_k = \sqrt{N_0^2 + (\epsilon^{-}_k -  {\bar H})^2}; ~
E^{-+}_k = \sqrt{N_0^2 + (\epsilon^{-}_k +  {\bar H})^2}$.
The self-consistency condition now takes the form
\begin{equation}
1 = \frac{U}{2}  {\sum_k}^{\prime} \left( \frac{n_{\beta_k} -
n_{\alpha_k}}{E^{--}_k} + \frac{n_{\delta_k} -
n_{\gamma_k}}{E^{-+}_k} \right) .
\label{self_cons_H}
\end{equation}
The computation of susceptibilities proceeds in the same way as without field,
the only new feature is the appearance of a cross-polarization
 term $\chi^{xz}_0$ which makes RPA ladder
summation a $2 \times 2$ matrix problem even if we
neglect, as we did earlier, small terms with a momentum transfer $Q$.
Specifically, we found
\begin{eqnarray}
\chi_{yy} (q,\omega) &=& \frac{\chi^{yy}_0 (q,\omega)
 (1 - 2U \chi^{zz}_0 (q,\omega)) - 2U |\chi^{zy}_0 (q,\omega)|^2}{(1 -
2U \chi^{yy}_0 (q,\omega))(1 - 2U \chi^{zz}_0 (q,\omega)) - 4U^2
|\chi^{zy}_0 (q,\omega)|^2} \nonumber \\
\chi_{yz} (q,\omega) &=& \frac{\chi^{yz}_0 (q,\omega)}{(1 -
2U \chi^{yy}_0 (q,\omega))
(1 - 2U \chi^{zz}_0 (q,\omega)) - 4U^2 |\chi^{zy}_0 (q,\omega)|^2} .
\label{chi_H}
\end{eqnarray}
Below, we restrict the consideration to an analysis of the
frequency dependence of the susceptibilities at $q =Q$.
 In this case,
the bare susceptibilities are the following
\begin{eqnarray}
\chi^{yy}_0 (Q,\omega) &=& \frac{1}{4N} {\sum_k}^{\prime}
(n_{\beta_k} -
n_{\alpha_k}) \left( \frac{1}{2{E^{--}_k} - \omega} + \frac{1}{2{E^{--}_k} +
 \omega} \right) + \nonumber \\
&&  (n_{\delta_k} -
n_{\gamma_k}) \left( \frac{1}{2{E^{-+}_k} - \omega} + \frac{1}{2{E^{-+}_k} +
 \omega} \right) \nonumber \\
\chi^{zz}_0 (Q,\omega) &=& \frac{1}{4N} {\sum_k}^{\prime}
(n_{\beta_k} -
n_{\alpha_k}) \left(\frac{\epsilon^{-}_k - \mu_B H}{E^{--}_k}\right)^2
\left( \frac{1}{2{E^{--}_k} - \omega} + \frac{1}{2{E^{--}_k} +
 \omega} \right) + \nonumber \\
&& (n_{\delta_k} -
n_{\gamma_k}) \left(\frac{\epsilon^{-}_k + \mu_B H}{E^{-+}_k}\right)^2
\left( \frac{1}{2{E^{-+}_k} - \omega} + \frac{1}{2{E^{-+}_k} +
 \omega} \right) \nonumber \\
\chi^{zy}_0 (Q,\omega) &=& i\frac{1}{4N} {\sum_k}^{\prime}
(n_{\beta_k} -
n_{\alpha_k}) \left(\frac{\epsilon^{-}_k - \mu_B H}{E^{--}_k}\right)
\left( \frac{1}{2{E^{--}_k} - \omega} - \frac{1}{2{E^{--}_k} +
 \omega} \right) - \nonumber \\
&& (n_{\delta_k} -
n_{\gamma_k}) \left(\frac{\epsilon^{-}_k + \mu_B H}{E^{-+}_k}\right)
\left( \frac{1}{2{E^{-+}_k} - \omega} - \frac{1}{2{E^{-+}_k} +
 \omega} \right) .
\label{chi_bar_h}
\end{eqnarray}
Performing the computations, we found that
\begin{eqnarray}
\chi^{yy}_0 (Q,\omega) &=& \frac{1}{2U} + c_1\frac{\omega^2}{t^3} \left(1 +
\frac{{\bar H}^2}{{\bar H}^{2}_0}\right),~~
 \chi^{zz}_0 (Q,\omega) = \chi^{yy}_0 (Q,\omega) -
c_2 N_0^2/t^3, \nonumber \\
&& \chi^{zy}_0 (Q,\omega) = i c_3 \omega {\bar H} /t^3
\label{chi_0H}
\end{eqnarray}
where  $c_i = c_i (t^{\prime}/t)$ and $H_0 \sim t$.
For the
type $A$ transition, we found that all three coefficients remain finite at
the critical point. Elementary considerations then show that the dominant
field dependence  in the
denominators of (\ref{chi_bar_h}) comes from $(\chi^{zy}_0)^2$ and is in the
form $\omega^2 H^2$. For the $n-$field action, this implies that the extra term
in the presence of $H$ has a form of Eq. (\ref{field}) (to see this, one simply
has to compute correlator $n^{z}n^{y} \propto \chi^{zy}$).
For a  disordering
transition of type $B$, the situation is more complex. The most relevant point
is that $c_3$ is still finite at the transition which implies that
there is no term $\sim sgn (\omega)$ in the coupling to the field.
However, we also found that $c_1$ and $c_2$ behave as
$c_1 = (1/2) D ((t/\sqrt{4N_0^2 - \omega^2}) +\ldots), c_2
= D ((t/2N_0)+\ldots)$
where $D$ was
defined after (\ref{chi_pi}) and dots stand for non-singular
(field-dependent) terms.
Right at the critical point,
$c_1 = [(i/2) D t/\omega + \ldots]$, and we have
$\chi^{yy,zz}_0 = (1/2U) - i\gamma\omega (1 -
{\bar H}^2/{\bar H}^{2}_0)/(4U^2) +
O(\omega^2 (1 - H^2/H^{2}_1))$.
 We see therefore that for $B-$ type transitions, $\gamma$
acquires a field-dependent part: $\gamma \rightarrow \gamma
 (1 - {\bar H}^2/{\bar H}^{2}_0)$.
Substituting this expression into (\ref{chi_bar_h}), we find
an extra field dependence in the denominator which has the same $\omega^2
{\bar H}^2$ form as the dependence imposed by  $(\chi^{zy}_0)^2$, but
with extra prefactor $(\gamma/t)^2$.
This implies that for $z=2$ transition, there are two different
relevant terms in the $n-$field action which describe the
coupling of the antiferromagnetic fluctuations
to the field. We discuss the relative importance of these terms in
Sec.~\ref{secunisus}. Finally, the effective action indeed contains a term
${\vec H}^2 - ({\vec H}{\vec n})^2 = ({\vec n}\times {\vec H})^2$ which
can be extracted from the longitudinal susceptibility $(\chi^{xx} (0,0))^{-1}
\propto H^2$. However, because of the constraint, this term contributes only
to uniform susceptibility at $T=0$ which is non-universal and is not considered
in this paper.

\begin{figure} \caption{Sketch of the imaginary part of the local dynamic
susceptibility,
$\chi_L^{\prime\prime}$
(obtained by integrating over momenta in the vicinity of the antiferromagnetic
ordering wavevector) of a nearly antiferromagnetic Fermi liquid in two
dimensions
at zero temperature.
({\em a\/}) The $T=0$ damping constant
$\Gamma$ is exactly 0,
as is the case for an insulating antiferromagnet in a quantum disordered
phase; $\Delta$ is
then the spin gap.
({\em b\/}) Shows the consequences of a small value of
$\Gamma$; there is now no
true gap, only a gap-like knee in the spectrum.
({\em c\/}) A large value of $\Gamma$ makes the spectrum relatively
featureless.
 }
\label{fig1}
\end{figure}

\begin{figure} \caption{({\em a\/}) Fermi surface of type $A$: ordering
wavevector $Q
=(\pi,\pi)$ cannot connect two points on the Fermi surface; ({\em b\/})
Fermi surface
of type $B$: $Q$ can connect two points on the Fermi surface. Dark regions
correspond to occupied electronic states in the first Brillouin zone.}
\label{fermi-surface}
\end{figure}

\begin{figure} \caption{The evolution of the Fermi surface with doping in the
Hubbard model with first and second neighbor hopping.
Shaded regions
correspond to hole pockets and black regions correspond to electron
pockets (i.e., doubly occupied states in the first Brillouin zone).
The description of the
figures is given in the text.}
\label{evol_fs}
\end{figure}

\begin{figure}
\caption{
Universal crossover behavior of the inverse antiferromagnetic correlation
length
$1/\xi$ (measured in units of $1/\hbar c$) as a function of the temperature
$T$, for a variety
of values of $\Gamma / \Delta$. The plot is the $N=\infty$ result contained
in Eqns
(\protect\ref{meq}) and (\protect\ref{xieq}). There are no arbitrary scale
factors,
and the scales of both axes are universal and meaningful.
}
\label{xifig}
\end{figure}

\begin{figure}
\caption{
As in Fig~\protect\ref{xifig}, but now with results for the
temperature-dependent part of
the uniform spin susceptibility $\chi_u (T) - \chi_u (0)$ measured in units of
$(g_G \mu_B /\hbar c)^2$. The non-universal constant $\alpha^{\prime}$ in
(\protect\ref{alphaprime}) has been set equal to 0.
The plot is the $N=\infty$ result contained in Eqns (\protect\ref{meq}) and
(\protect\ref{chieq}). Apart from the non-universal scale factor associated
with
$\alpha^{\prime}$, the scales of both axes are
universal and meaningful. }
\label{chifig}
\end{figure}

\begin{figure}
\caption{
Scaling regimes are sketched versus the normalized temperature,
$T/\Delta$, and the $T=0$ parameter $\Gamma/\Delta$
which describes type $A$ to type $B$
crossover. While the asymptotic $T\to 0$ behavior
is the same for all nonzero $\Gamma/\Delta$, an important
distinction between the
quantum disordered (QD), $z=1$ regime at small $\Gamma/\Delta$
and the QD, $z=2$ regime at large $\Gamma/\Delta$ is the presence of a
knee (see Fig~\protect\ref{fig1}b)
in $\chi''(\omega)$ at $\omega\sim \Delta$ in the former case.
We emphasize that the $T=0$ parameters $\Delta$ and $\Gamma$ may have a
highly nontrivial, possibly non-monotonic, doping dependence, therefore
the boundaries between different regimes on this phase diagram may have
somewhat different structure compared to
a phase diagram plotted as a function of doping.}
\label{sketch}
\end{figure}

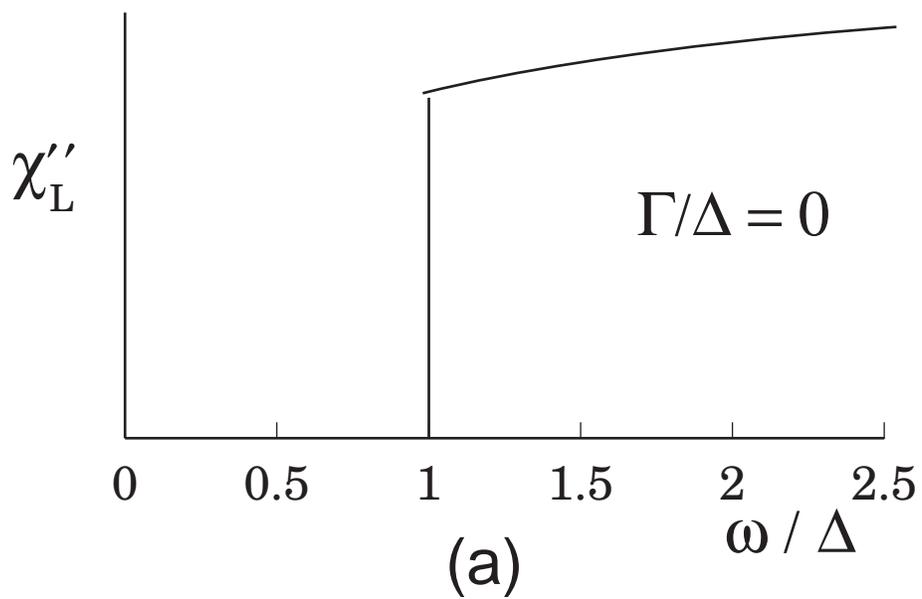
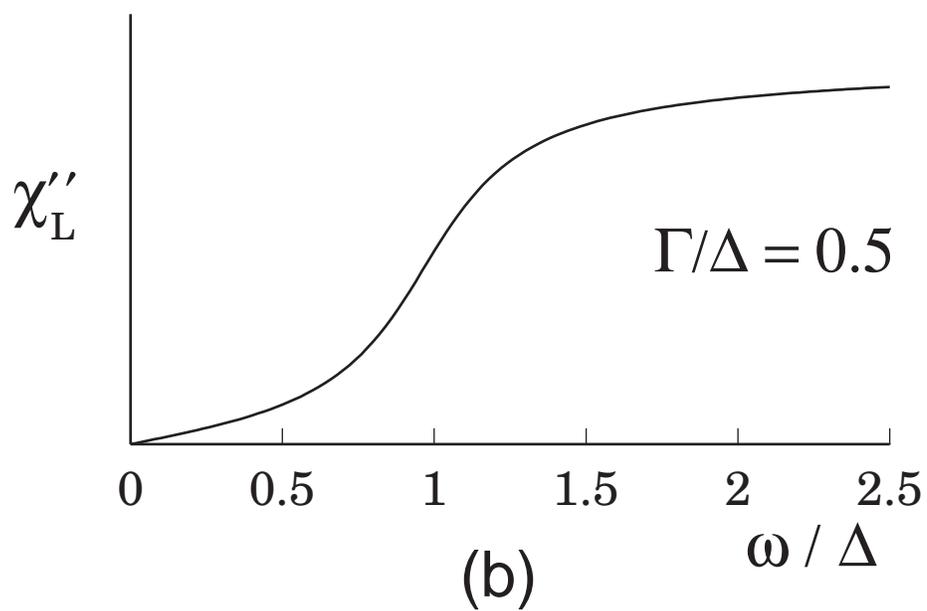
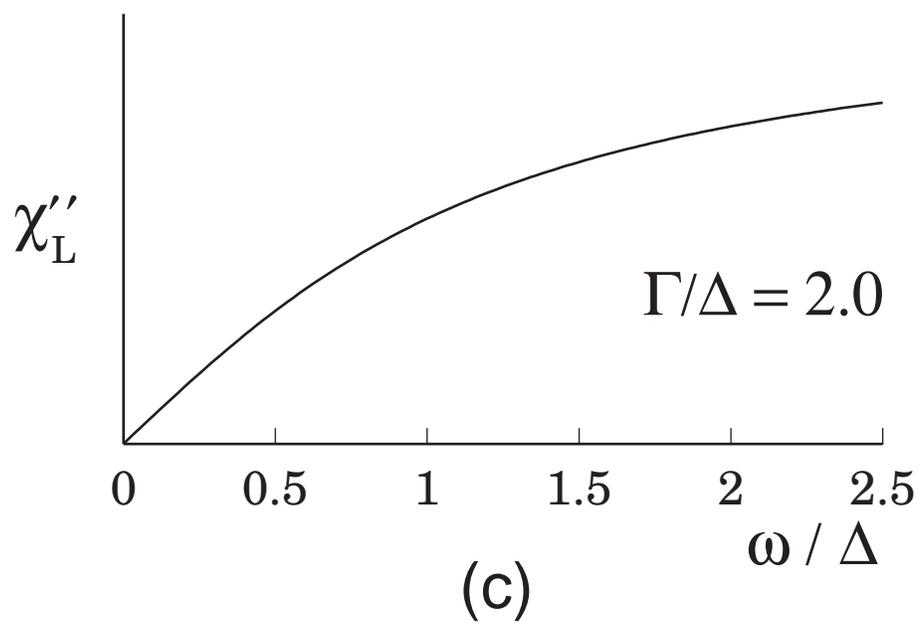
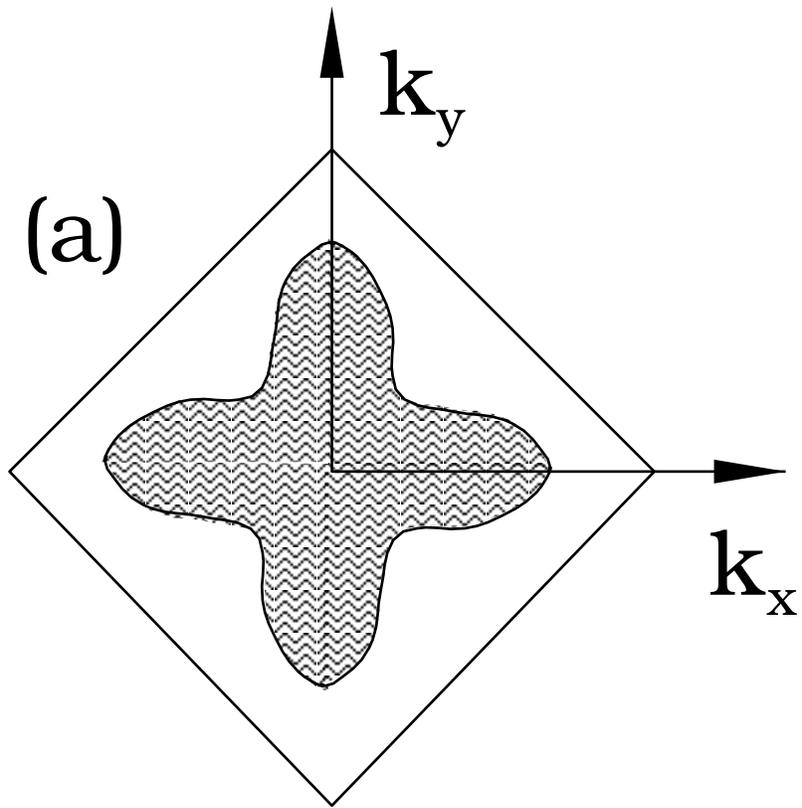
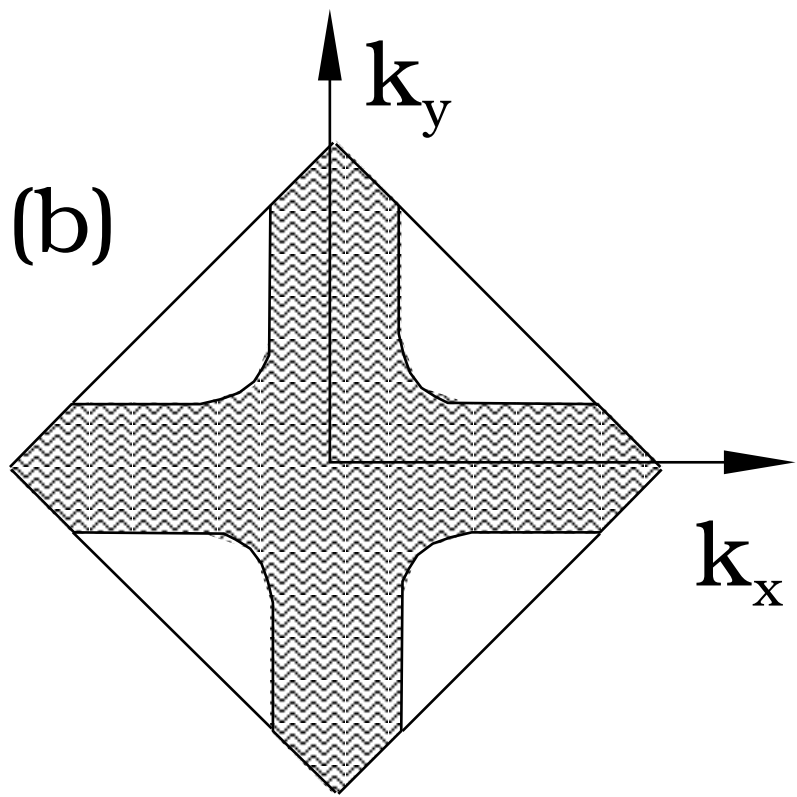
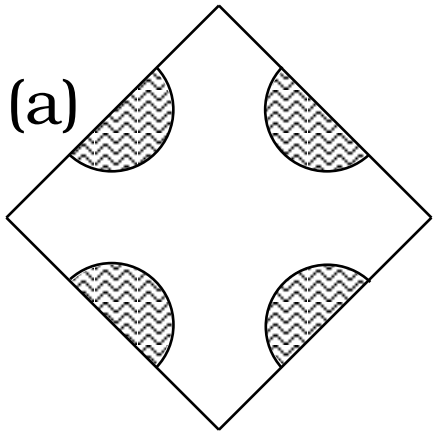
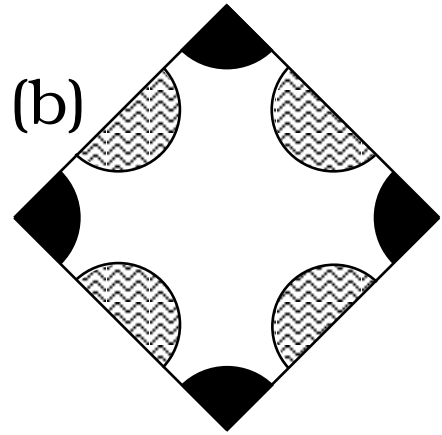
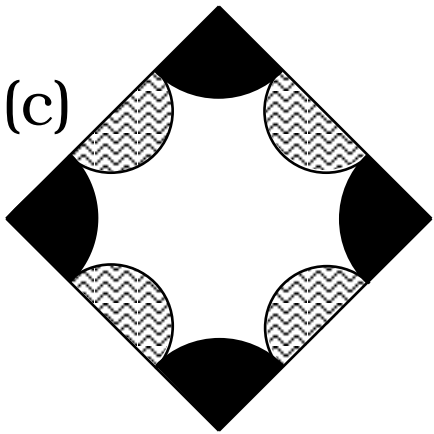
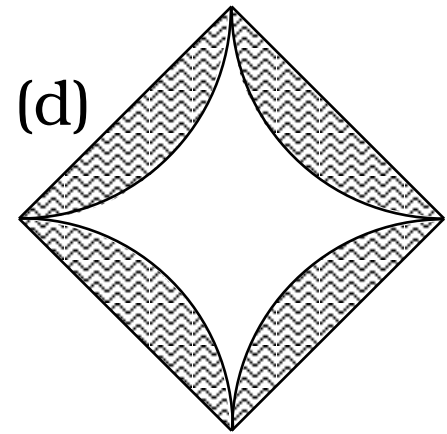
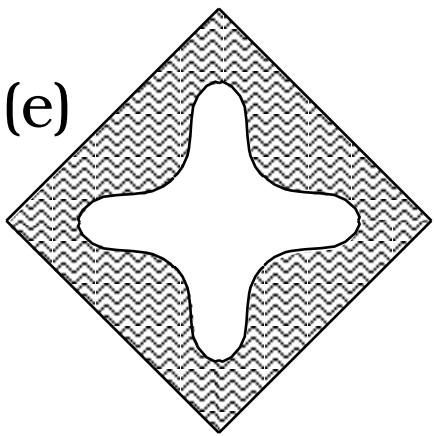
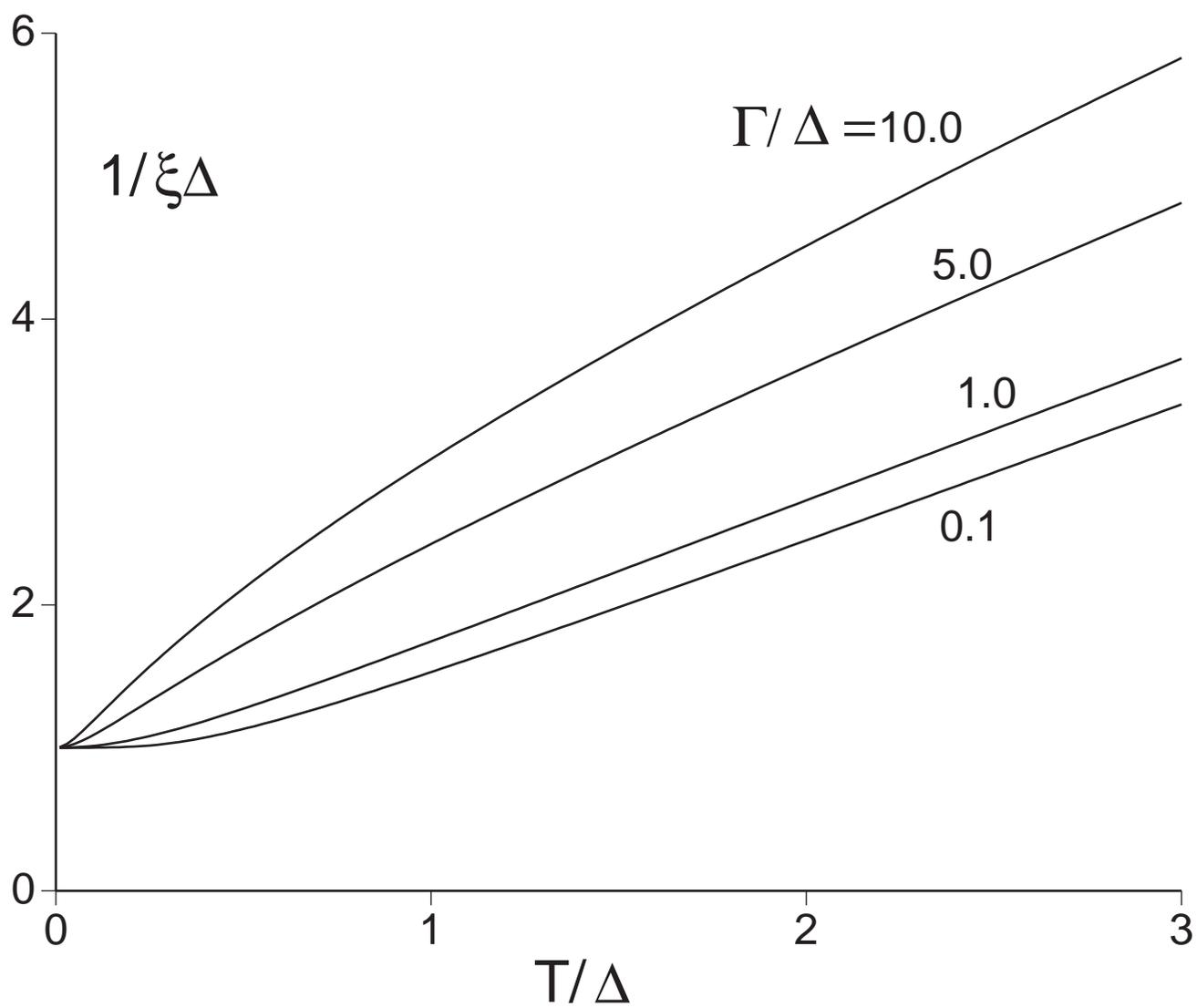
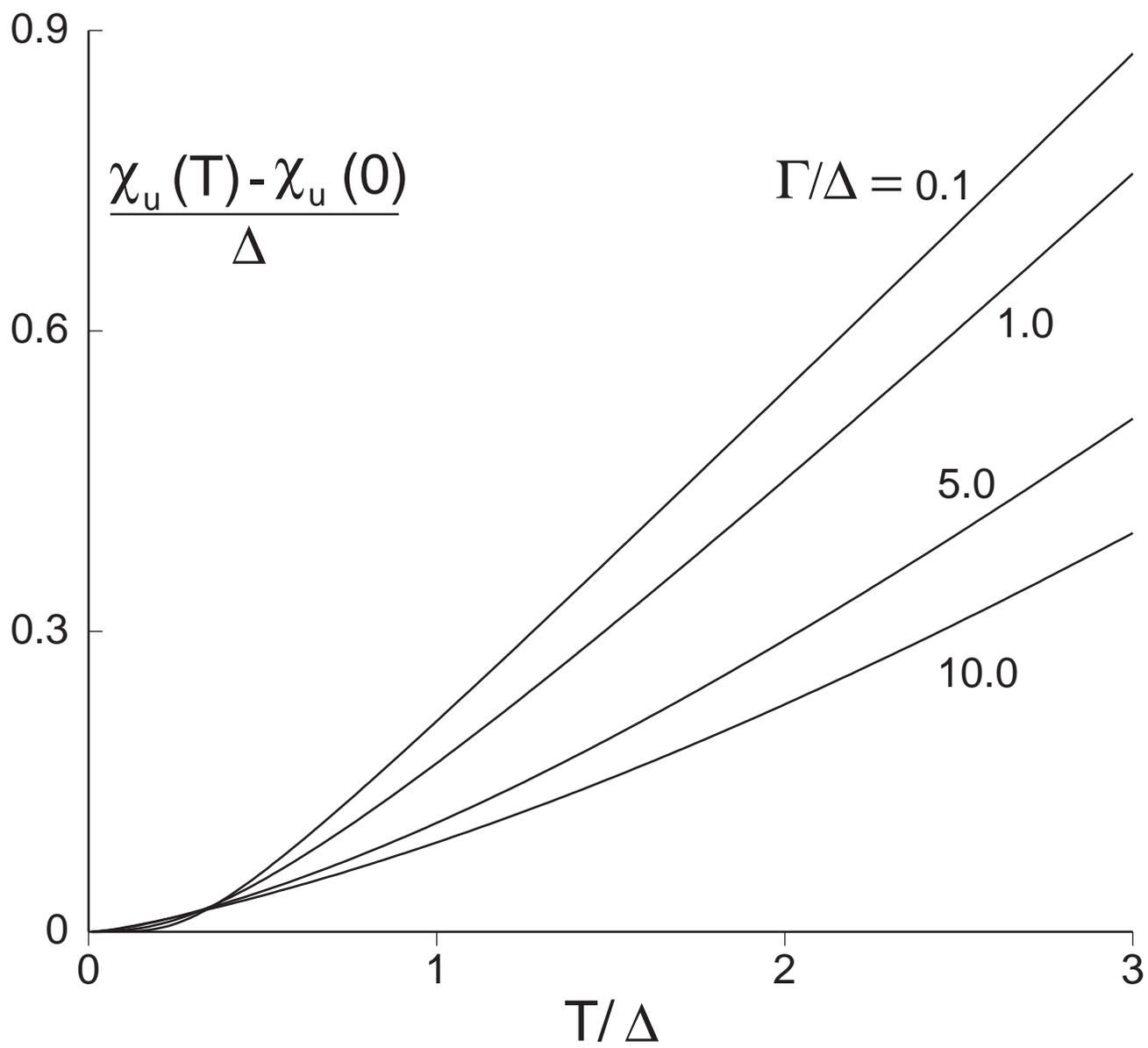
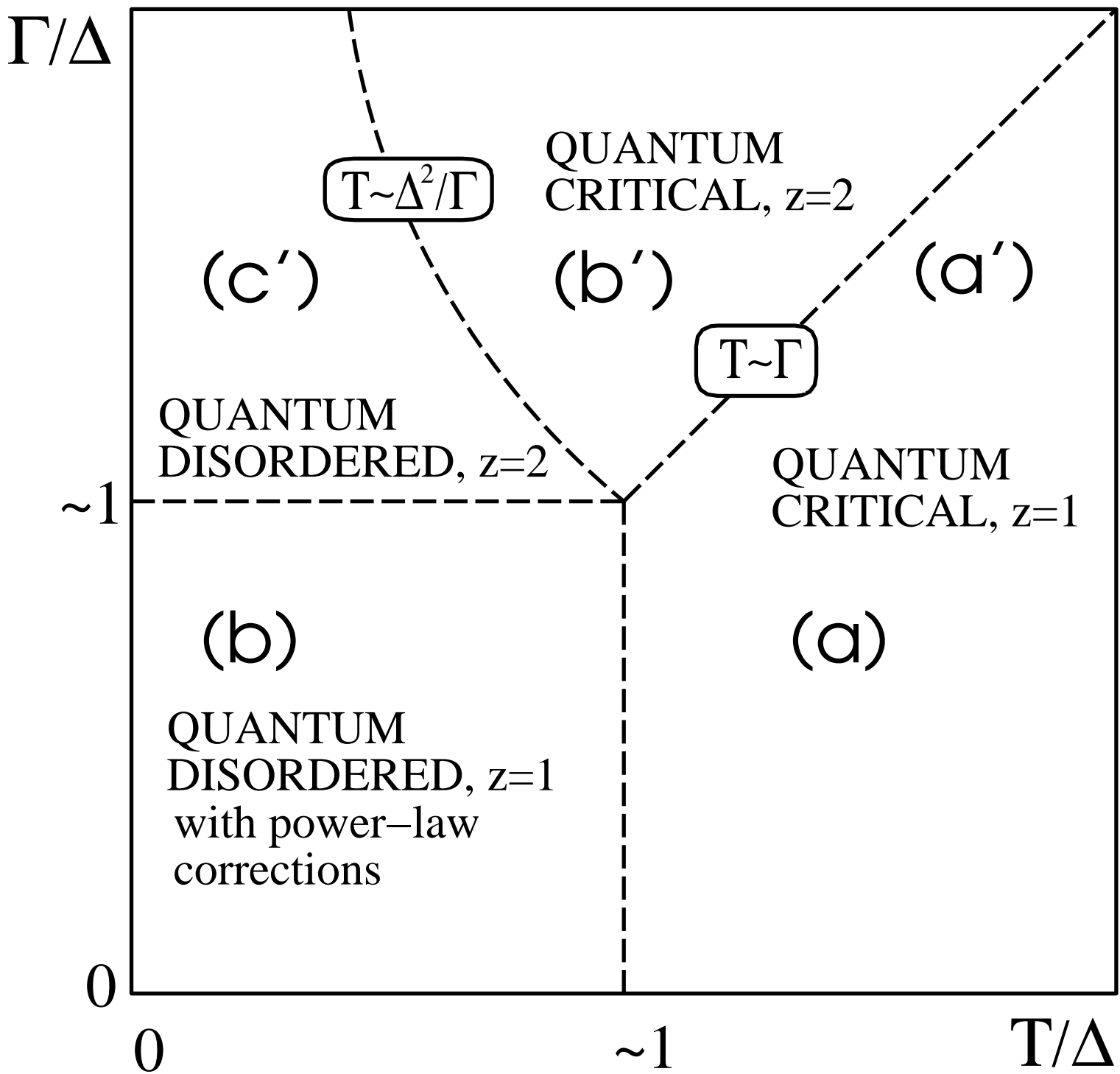
\end{document}